\documentclass[10pt,preprint]{aastex}               	% default manuscript style
\newcommand{\km}{${\rm km\,s}^{-1}$}
\newcommand{\fuse}{{\em FUSE}}
\newcommand{\hi}{H$\;${\small\rm I}\relax}

\newcommand{\neviii}{Ne$\;${\small\rm VIII}\relax}

\newcommand{\cii}{C$\;${\small\rm II}\relax}
\newcommand{\ciii}{C$\;${\small\rm III}\relax}
\newcommand{\civ}{C$\;${\small\rm IV}\relax}
\newcommand{\nni}{N$\;${\small\rm I}\relax}
\newcommand{\nii}{N$\;${\small\rm II}\relax}
\newcommand{\niii}{N$\;${\small\rm III}\relax}
\newcommand{\niv}{N$\;${\small\rm IV}\relax}
\newcommand{\nv}{N$\;${\small\rm V}\relax}
\newcommand{\oi}{O$\;${\small\rm I}\relax}
\newcommand{\oii}{O$\;${\small\rm II}\relax}
\newcommand{\oiii}{O$\;${\small\rm III}\relax}
\newcommand{\oiv}{O$\;${\small\rm IV}\relax}
\newcommand{\ov}{O$\;${\small\rm V}\relax}
\newcommand{\ovi}{O$\;${\small\rm VI}\relax}
\newcommand{\ovii}{O$\;${\small\rm VII}\relax}
\newcommand{\oviii}{O$\;${\small\rm VIII}\relax}
\newcommand{\pii}{P$\;${\small\rm II}\relax}
\newcommand{\sii}{S$\;${\small\rm II}\relax}
\newcommand{\siii}{Si$\;${\small\rm II}\relax}
\newcommand{\siiii}{Si$\;${\small\rm III}\relax}
\newcommand{\siiv}{Si$\;${\small\rm IV}\relax}
\newcommand{\siv}{S$\;${\small\rm IV}\relax}
\newcommand{\sv}{S$\;${\small\rm V}\relax}
\newcommand{\svi}{S$\;${\small\rm VI}\relax}
\newcommand{\feii}{Fe$\;${\small\rm II}\relax}

\slugcomment{Accepted for publication in the ApJS}
\shortauthors{Lehner et al.}
\shorttitle{Metal-Line Systems in the Low-$z$ Universe}
\begin{document}

\title{Low Redshift Intergalactic Absorption Lines in the Spectrum of HE\,0226--4110\altaffilmark{1}}

\author{N.\ Lehner\altaffilmark{2,3},
	B.\ D.\ Savage\altaffilmark{2},
	B.\ P.\ Wakker\altaffilmark{2},
	K.\ R. Sembach\altaffilmark{4},
	T.\ M. Tripp\altaffilmark{5}
	}
\altaffiltext{1}{Based on observations made with the NASA-CNES-CSA 
Far Ultraviolet Spectroscopic Explorer. FUSE is operated for NASA by the Johns 
Hopkins University under NASA contract NAS5-32985. Based on observations made with the NASA/ESA Hubble Space Telescope,
obtained at the Space Telescope Science Institute, which is operated by the
Association of Universities for Research in Astronomy, Inc. under NASA
contract No. NAS5-26555.}
\altaffiltext{2}{Department of Astronomy, University of Wisconsin, 475 North Charter Street, Madison, WI 53706}
\altaffiltext{3}{Department of Physics, University of Notre Dame, 225 Nieuwland Science Hall, Notre Dame, IN 46556}
\altaffiltext{4}{Space Telescope Science Institute, 3700 San Martin Drive, Baltimore, MD 21218.}
\altaffiltext{5}{Department of Astronomy, University of Massachusetts, Amherst, MA 01003.}

\begin{abstract}
We present an  analysis of the {\em Far Ultraviolet Spectroscopic Explorer (FUSE)}
and the Space Telescope  Imaging Spectrograph (STIS E140M) 
spectra of HE\,0226--4110 ($z_{\rm em} = 0.495$) that have a nearly continuous wavelength coverage 
from 910 to 1730 \AA.  We detect 56 Lyman absorbers and 5 \ovi\ absorbers.  
The number of  intervening \ovi\ systems per unit redshift with $W_\lambda \ga 50$ m\AA\ 
is $d{\mathcal N}({\mbox \ovi})/dz \approx 11$.
For 4 of the 5 \ovi\ systems other ions (such as \ciii, \civ, \oiii, \oiv) are detected. 
The \ovi\ systems unambiguously trace  hot gas only in one case. 
For the 4 other \ovi\ systems, photoionization  and collisional ionization models are viable options
to explain the observed column densities of the \ovi\ and the other ions.  
If photoionization applies for those systems, 
the broadening of the metal lines must be mostly non-thermal or several components 
may be hidden in the noise, but the \hi\ broadening appears to be mostly thermal. 
If the \ovi\ systems are mostly photoionized, only a fraction of the observed \ovi\ will contribute to 
the baryonic density of the  warm-hot ionized medium (WHIM) along this line of sight. 
Combining our results with previous ones, we show that
there is a general increase of $N$(\ovi) with increasing $b$(\ovi). 
Cooling flow models can reproduce the $N$--$b$ distribution but fail to reproduce the 
observed ionic ratios. 
A comparison of  the number of \oi, \oii, \oiii, \oiv, and \ovi\ systems per unit redshift 
show that the low-$z$ IGM  is more highly ionized than weakly ionized. 
We confirm that photoionized \ovi\ systems show a decreasing ionization parameter
with increasing \hi\ column density. \ovi\ absorbers 
with collisional ionization/photoionization degeneracy follow this relation, 
possibly suggesting that they are principally photoionized. 
We find that the photoionized \ovi\ systems in the low redshift IGM have a median abundance 
of 0.3 solar.  We do not find additional \neviii\  systems other than the one found by Savage et al.,
although our sensitivity should have allowed the detection  of \neviii\ in \ovi\ systems
at $T \sim (0.6-1.3 )\times 10^6$ K (if collisional ionization equilibrium applies).  
Since the bulk of the WHIM is believed to be at temperatures $T> 10^6 $ K,
the hot part of the WHIM remains to be discovered with FUV--EUV metal-line transitions.
\end{abstract}


\section{Introduction}
From the earliest time to the present-day epoch, most of the baryons are
found in the intergalactic medium (IGM). The Ly$\alpha$ forest is a
signature of the  IGM that is imprinted on the spectra of QSOs and allows
measurements of the evolution 
of the universe over a wide range of redshift. At $z > 1.5$, \hi\ in 
the IGM can be observed with ground-based 8--10 m telescopes. 
In the low redshift universe ($z\le 1.5$), the Ly$\alpha$ transition and most of the metal resonance
lines fall in the ultraviolet (UV) wavelength range, requiring challenging space-based observations
of QSOs. Most Ly$\alpha$ absorbers at $z <1.5$  do not show detectable
metal absorption lines in currently available spectra, but those that
do show associated metals provide further information about the
abundances, kinematics, and ionization corrections in the low-redshift
IGM and halos of nearby galaxies.  Metal-line
systems also give indications about the metallicity evolution of
the IGM and can contain large reservoirs of baryons at
low and high redshift \citep[e.g.,][]{tripp00b,carswell02,bergeron05}.
Detecting  metal-line systems in the IGM requires not only high signal-to-noise
but also high spectral resolution FUV spectra and simple lines of sight 
to avoid blending between the different galactic and intergalactic absorbers. 
In the last few years, a few high quality spectra
of  QSOs covering the wavelength range from the Lyman limit to about 1730 \AA\ have
been obtained with the Space Telescope Imaging Spectrograph (STIS)
onboard the {\em Hubble Space Telescope (HST)}\ and {\em Far Ultraviolet Spectroscopic 
Explorer (FUSE)}, thereby  allowing sensitive measurements of the metal-line systems in the 
tenuous IGM \citep[e.g.,][]{tripp01,savage02,prochaska04,richter04,sembach04,williger05}.

The cold dark matter cosmological simulations provide
a self-consistent explanation of the Ly$\alpha$ absorbers seen in the QSO 
spectra \citep[e.g.,][]{dave99,dave01}. At high redshift, they predict
that most of the Ly$\alpha$ absorbers consist of cool (less than a few $10^4$ K) 
photoionized gas and virtually all baryons are observed in this gas-phase at $z>2$ 
\citep[e.g.,][]{weinberg97,rauch97}.
As the universe expands, the initial density perturbations collapse, producing shock-heated gas 
at temperatures of $10^5$--$10^7$ K \citep{cen99,dave99,fang02}. At the present epoch
($z\la 1 $), the hydrodynamical simulations predict that $\sim$30--50\% of the normal 
baryonic matter of the universe lies in a tenuous warm-hot intergalactic medium (WHIM), and another 
$\sim$30\% of the baryons lies in a cooler, photoionized, tenuous intergalactic 
gas; observations appear to support the prediction regarding low-$z$ photoionized gas
\citep[e.g.,][]{penton04}. 

At temperatures of $10^5$--$10^7$ K, the most abundant
elements C, N, O, and Ne are highly ionized.  Detecting the hot component
of the WHIM at $T>10^6$ K is possible through measurements of X-ray \ion{O}{7} K$\alpha$
and \ion{O}{8} absorption lines at 21.870 and 18.973+18.605 \AA,
respectively. These have been detected at low redshift with {\it Chandra}
along one line of sight \citep{nicastro05}. However, apart from detections at $z=0$, most of the 
current X-ray observations of \ovii\ and \oviii\ reported  in the IGM remain marginal 
or the claims are contradicted by higher signal-to-noise spectra \citep{rasmussen03}.
The EUV lines of \neviii\ $\lambda$$\lambda$770, 780 tracing gas at $7\times 10^5$ K
are redshifted to the ultraviolet (UV) wavelength region for $z>0.18$ and \citet{savage05} reported
the first detection of the \neviii\ doublet in the spectrum of the bright QSO HE\,0226--4110. 
The cooler part ($T\la 5 \times 10^5$ K) of 
the WHIM is currently better observed via the \ovi\ doublet at 1031.926 and 1037.617 \AA,
which can be efficiently observed by combining STIS and {\fuse}\ observations 
\citep[e.g.,][]{tripp00a,tripp00b,tripp04,danforth05}. 
However, the \ovi\ absorbers can  arise in warm collisionally ionized gas ($T\sim 3 \times 10^5$ K),
as well as in cooler, photoionized, low density gas
\citep[e.g.,][]{tripp01,savage02,prochaska04}. A study of the origin(s)
of ionization of the \ovi\ is therefore important for estimates of the density 
of baryons in the WHIM and the IGM in general.

In this paper, we report the complete {\fuse}\ and STIS observations 
and analyses of the IGM absorption lines in the spectrum of the 
$z=0.495$ quasar HE\,0226--4110. From these observations we 
derive accurate cloud parameters (redshift, 
column density, Doppler width) for clouds in the Lyman forest and any associated metals. 
While all the  IGM measurements for HE\,0226--4110 are reported here, the 
metal-free Ly$\alpha$ forest observations will be discussed in more detail
in a future paper where we will combine the 
present results with results from other sight lines in order to derive the physical and statistical 
properties of the  Ly$\alpha$ forest at low redshift.  
We concentrate in this work on defining the origin(s) of 
the observed \ovi\ systems and the implications for the
IGM by combining our observations with recent analyses of other QSO spectra
at low redshift \citep[in particular,][]{tripp01,prochaska04,richter04,sembach04,williger05}.

The physical properties and ionization conditions in metal-line 
absorbers can be most effectively studied by obtaining observations of 
several species in different ionization stages.
Using the same species in several ionizing stages provides a direct way
to constrain simultaneously metallicity, ionization, and physical conditions. For example, 
for oxygen, EUV measurements of \oii, \oiii, \oiv, \ov\ 
can be combined with FUV measurements of \oi\
and \ovi. The HE\,0226--4110 line of sight is particularly favorable for a search for EUV and FUV absorptions
because the H$_2$ absorption in our Galaxy is very weak,
producing a relatively clean FUV spectrum for IGM studies. 
In particular, we systematically search for and report the measurements of 
\ciii, \oiii, \oiv, \ovi, and \neviii\ associated with each Ly$\alpha$ absorber
in order to better characterize the physical states of the Ly$\alpha$ forest
at low redshift. 

The organization of this paper is as follows. After describing 
the observations and data reduction of the HE\,0226--4110 spectrum in \S\ref{obsred}, we 
present the line identification and analysis to estimate the redshift, column density ($N$),
Doppler parameter ($b$) for the IGM clouds in \S\ref{anal}. In \S\ref{review}, we determine
the physical properties and abundances of the metal-line absorbers
observed toward HE\,0226--4110. The implications of our results combined 
with recent observations of the IGM are discussed in \S\ref{discuss}. 
A summary of the main results is presented in \S\ref{sum}.

\section{Observations and Data Reduction}\label{obsred}
\begin{figure}[tbp]
\epsscale{1} 
\plotone{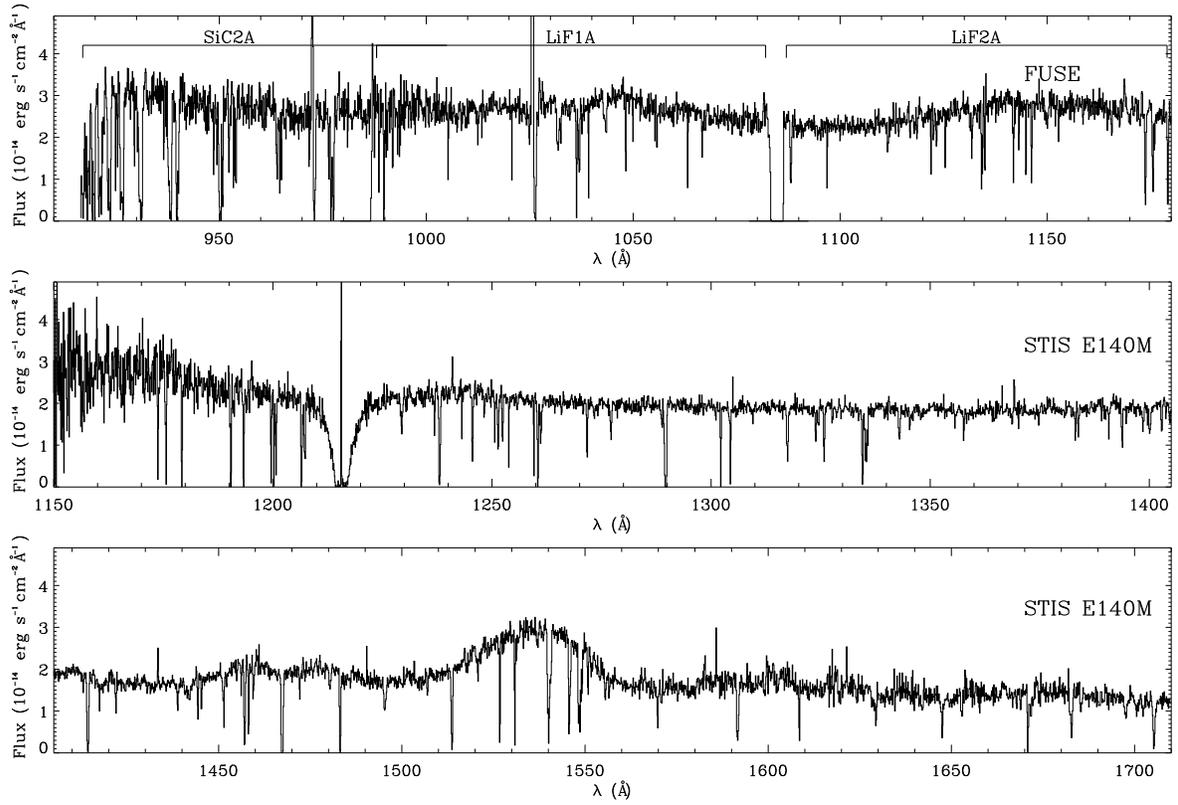}
\caption{{\em FUSE}\ and STIS FUV spectra of HE\,0226--4110.
The data have been binned into 0.1 \AA\ samples and 
we only present the {\em FUSE}\ night data and for clarity we only present the {\em FUSE}\ night data in this
illustration. The spike in the middle of the damped Galactic Ly$\alpha$ profile is
the geocoronal Ly$\alpha$ emission from the Earth's atmosphere. \label{fig1}}
\end{figure}

We have obtained a high quality FUV spectrum of 
HE\,0226--4110 [$(l,b) = (253\fdg94, -65\fdg78$); $z_{\rm em} = 0.495$]
 covering  nearly continuously 
the wavelength range from 916 to 1730 \AA. 
To show the broad ultraviolet continuum shape of HE0226-4110 as well
as the locations and shapes of the QSO emission lines, we display in
Figure~\ref{fig1} the {\it FUSE}\ and STIS data with a spectral bin size of 0.1
\AA.  This figure shows the overall quality of the spectrum and the
spectral regions covered by the various {\it FUSE}\ channels and the
STIS spectrum. Several emission features can be discerned, which are
associated with gas near the QSO (see Ganguly et al. 2005, in
preparation). The feature near 1537~\AA\ is mostly Ly$\beta$ emission, with
possibly some \ovi\ $\lambda$1031 and \ovi $\lambda$1037 emission mixed in.
Around 1454~\AA, weak Ly$\gamma$ emission is visible. Intrinsic \neviii\
emission may be present in the 1150-1160~\AA\ region. The feature centered
on 1049~\AA\ is most likely \oiii\ $\lambda$702.332 emission. A corresponding,
slightly weaker \oiii\ $\lambda$832.927 emission feature can be seen near
1240~\AA. Figure~\ref{fig1} also shows that strong lines are sparse in this
sight line; the weakness of Galactic H$_{2}$ lines in particular makes
this a valuable sight line for the study of extragalactic EUV/FUV
absorption lines. We note that occasional
hot pixels in the STIS spectrum are clearly
evident in this figure, and the possible contamination of absorption
profiles by warm/hot pixels must be borne in mind. In
addition, residual geocoronal Ly$\alpha$, Ly$\beta$ and Ly$\gamma$ emission is visible. 
All wavelengths and velocities are given 
in the heliocentric reference frame in this 
paper. Toward HE\,0226--4110, the Local Standard
of Rest (LSR) and heliocentric reference frames 
are related by $v_{\rm LSR} = v_{\rm helio}-14.3$ \km. 
\citet{savage05} described in detail the data used 
in this paper and its processing, and we give below only a brief summary.

\subsection{{\em HST}/STIS Observations}\label{stisobs}
The STIS observations of HE\,0226--4110 (GO program 9184, PI: Tripp) were obtained with the
E140M intermediate-resolution echelle grating between 2002 December
26 and 2003 January 1, with a total integration time
of 43.5 ksec (see Table~\ref{t1a}).
The entrance slit was set to $0\farcs2\times 0\farcs06$. 
The spectral resolution is 7 \km\ with a detector pixel size 
of 3.5 \km. The S/N per 7 \km\ resolution element of the 
spectrum of HE\,0226--4110 is 11, 11, and 8, at 1250, 1500, 
1600 \AA, respectively. The S/N is substantially lower for 
$\lambda\la 1180$ \AA\ and $\lambda \ga 1650$ \AA.

The STIS data reductions provide an excellent wavelength calibration.
The velocity uncertainty is $\sim$1 \km\ with occasional errors
as large as 3 \km\ \citep[see the appendix of][]{tripp05}. 
In order to align the {\em FUSE}\ spectra that have an uncertain
absolute wavelength calibration (see \S~\ref{fuseobs}) with the 
STIS spectrum, we systematically measured the velocity of the 
interstellar lines that are free of blends.
Those lines are: \sii\ $\lambda$$\lambda$1250, 1253, 1259, 
\nni\ $\lambda$$\lambda$1199, 1200, 1201, \oi\ $\lambda$1302, 
\siii\ $\lambda$$\lambda$1190, 1193, 1304, 1526, \feii\ $\lambda$1608,
\ion{Ni}{2} $\lambda$1370, and \ion{Al}{2} $\lambda$1670.
Using these species, we find $\langle v_{\rm ISM} \rangle_{\rm helio} = 10.4 \pm 1.4 $ \km\ toward HE\,0226--4110. 
We use this velocity to establish the zero point wavelength calibration 
of the {\em FUSE}\ observations.

\subsection{{\em FUSE} Observations}\label{fuseobs}

The {\em FUSE}\ observations of HE\,0226--4110 were  obtained between 
2000 and 2001 from the science team \ovi\ project \citep{wakker03,savage03,sembach03}, 
and between 2002 and 2003 as part of the {\fuse}\  GO program D027 (PI: Savage) 
(see Table~\ref{t1}). The total exposure time of these programs in  
segments 1A and 1B is 194 ks, in segments 2A and 2B 191 ks. The night data typically account 
for 65\% of the total exposure time.  
The measurements were obtained in time-tagged mode and cover the wavelengths 
between 916 to 1188 \AA\ with a spectral resolution of $\sim$20 \km. 
In the wavelength range 916 to 987 \AA\ only SiC\,2A data are available  because of channel
alignment problems. Over the 
wavelength region from 987 to 1182 \AA\ we used LiF\,1A, and from 1087 to
1182 \AA\ LiF\,2A was used. The lower S/N observations in LiF\,2B and LiF\,1B were used to 
check for fixed pattern noise in  LiF\,1A  and LiF\,2A observations, respectively. 
To reduce the effects of detector fixed-pattern noise,
some of the exposures were acquired using focal plane split motions (see Table~\ref{t1}), wherein subsequent
exposures are placed at different locations on the detector.

The spectra were processed with CALFUSE v2.1.6 or v2.4.0 (see Table~\ref{t1}). The most difficult task
was to bring the different extracted exposures into 
a common heliocentric reference frame before coadding them. To do so, 
we fitted the ISM lines in each segment of each of the 8
exposures and shifted them to the heliocentric frame measured with the STIS spectrum.  
Typically, in LiF\,1A, we use \ion{Si}{2} $\lambda$1020, \ion{Ar}{1} $\lambda$$\lambda$1048, 1066, 
\ion{Fe}{2} $\lambda$1063; in LiF\,1B/2A, \ion{Fe}{2} $\lambda$$\lambda$1096,
1112, 1121, 1125, 1142, 1143, 1144; in SiC\,2B \ion{Ar}{1} $\lambda$$\lambda$1048, 1066, 
\ion{Fe}{2} $\lambda$$\lambda$1063, 1096; and in SiC\,2A
\ion{O}{1} $\lambda$$\lambda$921, 924, 925, 929, 930, 936, 948, 950, 971, 976.
We forced the ISM lines in each exposure of the {\em FUSE}\ 
band to have $v_{\rm helio} = 10.4$ \km. 
The rms of the measured velocities of the ISM lines is typically 4--6 \km\ for 
the short exposures, 3--4 \km\ for the two longer ones. We therefore estimate 
that the velocity zero-point uncertainty for the {\em FUSE}\ data is $\sim$5 \km\ ($1\sigma$). The relative
velocity uncertainty is also $\sim$5 \km\ although it may be larger near the edge of 
the detector. 

The oversampled {\fuse}\ spectra were binned to a bin size of 4 pixels (0.027 \AA), providing
about three samples per 20 \km\ resolution element. Data taken during orbital day and orbital night were combined, 
except in cases where an airglow line contaminated a spectral region of interest. Then only
night data were employed. 
The S/N per 20 \km\ resolution element is typically 11 in SiC\,2A ($\lambda < 987$ \AA), and 18 
in LiF\,1A and LiF\,2A ($\lambda > 987$ \AA). 

\section{Analysis}\label{anal}
\subsection{Line Identification, Continuum, and Equivalent Width}

\begin{figure}[tbp]
\epsscale{0.9} 
\plotone{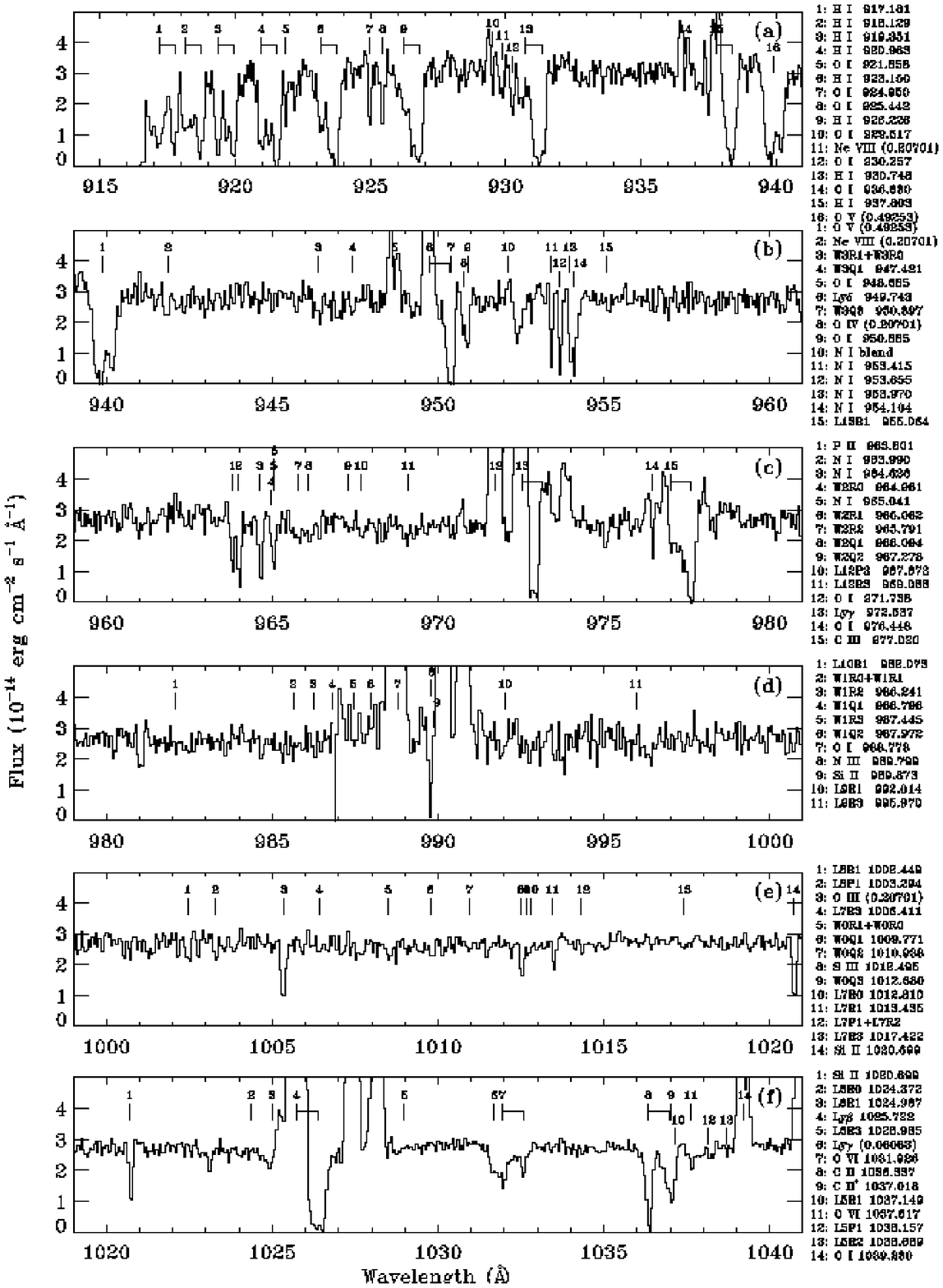}
\caption{\small {\fuse}\ spectra of HE\,0226--4110 as a function 
of the heliocentric wavelength between 917 and 1183 \AA. The 
data are binned by 4 pixels providing about three samples per 20 \km\ resolution element.
At $\lambda \la 986$ \AA, data are from SiC\,2A channel, between 
986 \AA\ and 1083 \AA, data are from LiF\,1A channel, and 
at $\lambda \ga 1086$ \AA\ data are from LiF\,2A channel. Line 
detections are denoted by tick marks above the spectra. Line
identifications for these detections are listed on the right-hand
side of each panel. Redshifts ({\em in parentheses}) are indicated 
for intergalactic lines. In cases where the $+190$ \km\ high velocity
interstellar feature is present, an offset tick mark is attached to 
the primary tick mark at the rest wavelength of the line. 
\label{fig2}}
\end{figure}

\begin{figure}[tbp]
\epsscale{0.9} 
\plotone{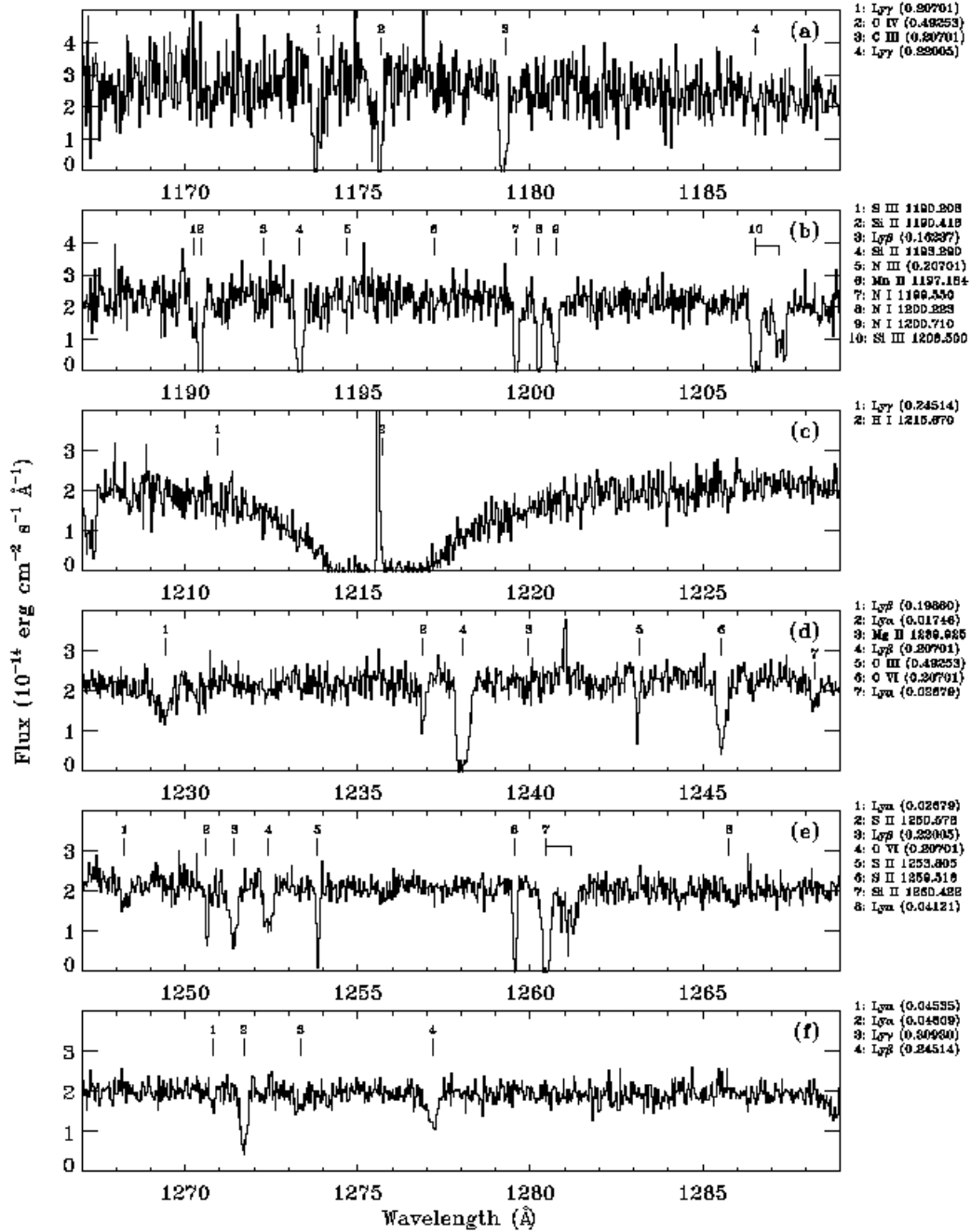}
\caption{\small {\em HST}\ STIS E140M spectra of HE\,0226--4110 as a function 
of the heliocentric wavelength between 1167 and 1730 \AA. The 
data have spectral resolution of about 7 \km. We have binned
the data into two pixels ($\approx 7$ \km) for clarity, but all the measurements were conducted
on the fully sampled data. Labels are similar to those in Fig.~\ref{fig2}.
\label{fig3}}
\end{figure}

We show in Figs.~\ref{fig2} and \ref{fig3} the {\fuse}\ and STIS spectra of HE\,0226--4110, 
respectively, where about 250 absorption features are identified. 
 All the ISM and IGM absorption
lines are labeled in Figs.~\ref{fig2} and \ref{fig3}.
We first identified  all the absorption features associated with interstellar resonance and excited 
UV absorption lines using the compilation of atomic parameters by \citet{morton03} and
the H$_2$ molecular
line list of \citet{abgrall93a,abgrall93b}. The EUV atomic parameters were obtained from 
\citet{verner96}.  Because HE\,0226--4110 is at high latitude
($b=-65\fdg78$) and in a favorable Galactic direction, 
the molecular absorption from H$_2$  remains very weak and greatly reduces the 
problem of blending with IGM absorptions. We identified every H$_2$ line in the HE\,0226--4110
{\fuse}\ spectrum and we modeled the H$_2$ lines by measuring the equivalent width in each 
$J= 0-4$ rotational level (see Wakker 2005). We found that the total  H$_2$ column density is $\log N({\rm H_2}) = 14.54$, 
corresponding to a very small molecular fraction of $f({\rm H_2}) = 3.7 \times 10^{-6}$. 
The atomic-ionic ISM gas component consists principally of two main 
clouds, a low-velocity component at $v_{\rm helio} = 10$ \km\ and a high-velocity component (HVC) at 190 \km
(see Fox et al. 2005, B. D. Savage et al. 2006, in prep.). There is also a weaker ISM absorber at about $-20$ \km\
(B. D. Savage et al. 2006, in prep.).
The HVC is detected in the high ions (\ion{O}{6}, \ion{Si}{4}, and \ion{C}{4}) and only in 
the strongest transitions of the low-ions (\ion{C}{2}, \ion{C}{3}, and \ion{Si}{2} 
$\lambda$1260) (see Fox et al. 2005, and see Figs.~\ref{fig2} and \ref{fig3}). 
Each of these velocity components has to be considered carefully for possible blending 
with IGM absorption. An example of such blending occurs between Ly$\alpha$ at $z= 0.08735$ 
and \ion{Ni}{2} $\lambda$1317.217. In the footnote of Table~\ref{t2}, 
we highlight any blending problems between ISM and IGM lines, and between IGM lines at different
redshifts. After identifying all the intervening Galactic absorption lines, we 
searched for Ly$\alpha$ absorption at $z>0$. For each  Ly$\alpha$ absorption line,
we checked for additional Lyman-series and associated metal lines. Since this line of sight
has so little H$_2$, it provides an unique opportunity to search for weak IGM metal lines. 
We systematically searched for the FUV lines of \ion{C}{3} $\lambda$977 and  the 
\ion{O}{6} $\lambda$$\lambda$1031, 1037 doublet, and when the redshift allows, for the EUV lines 
\ion{O}{3} $\lambda$832, \ion{O}{4} $\lambda$787, and the \ion{Ne}{8} $\lambda$$\lambda$770,780 doublet.
If one of these lines was lost in a terrestrial airglow emission line, we considered the night data only. 
Since the wavelength coverage is not complete to the redshift of the QSO and because shock
heated gas does not have to be associated with a narrow \hi\ system, we always made sure
that none of the Ly$\alpha$ systems was an \ovi\ or \neviii\ system by using the atomic properties
of these doublets. We found one possible \ovi\ system at 
$z=0.42663$ not associated with \hi\ Ly$\beta$ (Ly$\alpha$ being beyond detection at this
redshift). Note that we identify in Figs.~\ref{fig2} and \ref{fig3}
the associated system to the QSO at $z= 0.49253$, but we do not report any measurements 
(see Ganguly et al. 2005 for an analysis of this system, which is
associated with gas very close to the AGN). 
We find a total of 59 systems (excluding the associated system at $z= 0.49253$) toward
HE\,0226--4110. Two of the systems ($z=0.20701,0.27155$) clearly have multi-component \hi\ 
absorption (see \S~\ref{coldes}), and one possible system is detected only in \ovi\ at $z=0.42663$
(see also \S\ref{misid}). 

All our detection limits reported in this work (except otherwise stated) are 
3$\sigma$. The 3$\sigma$ limit will vary depending on its wavelength position 
in the spectrum because the S/N varies with wavelength.   
In Table~\ref{t2}, we report our measurements of the line strength
(equivalent width), line width (Doppler parameter), and column density
for all the detected  IGM species or the $3\sigma$ limits on the equivalent width and column density
for the non-detected species.

 All our measurements are in the rest-frame. 
The equivalent widths and uncertainties were measured following
the procedures of \citet{sembach92}. The adopted uncertainties for the derived equivalent widths,
column densities and Doppler parameters (see \S\ref{coldes}) are  $\pm 1\sigma$.  
These errors include the effects of statistical noise, fixed-pattern
noise for {\em FUSE}\ data when two or more channels were present, 
the systematic uncertainties of the continuum placement, and the  velocity range 
over which the absorption lines were integrated.  
The continuum levels were obtained by fitting low-order 
($<4$) Legendre polynomials within 500 to 1000 \km\ of each absorption line. 
For weak lines, several continuum placements 
were tested to be certain that the continuum error was robust. For 
the {\fuse} data we considered data from multiple channels whenever possible
to assess the fixed-pattern noise. 
An obvious strong fixed-pattern detector feature is present in the LiF\,1A channel
at 1043.45 \AA\ (see Fig~\ref{fig2}).

\subsection{Redshift, Column Density, and Doppler Parameter}\label{coldes}

To measure the centroids of the absorption line, the column densities
and the Doppler parameters, we systematically used two methods: the apparent optical depth 
(AOD, see Savage \& Sembach 1991) and a profile fitting method. 
To derive the column density and measure the redshift, we used the atomic parameters for the FUV  and 
EUV lines listed in \citet{morton03} and \citet{verner96}, respectively. 
Note that the system's redshifts in this paper refers to the \hi\ centroids, except for 
the \ovi\ system at $z=0.42660$ 

In the AOD method, the absorption profiles are converted into apparent optical depth (AOD) per unit velocity,
$\tau_a(v) = \ln[I_{\rm c}/I_{\rm obs}(v)]$, where $I_{\rm obs}$,
$I_{\rm c}$ are the intensity with and without the absorption,
respectively.  The AOD, $\tau_a(v)$, is related to the apparent column
density per unit velocity, $N_a(v)$, through the relation
$N_a(v) = 3.768 \times 10^{14} \tau_a(v)/(f \lambda(\mbox{\AA})$)  ${\rm cm}^{-2}\,({\rm km\,s^{-1}})^{-1}$.
The total column density is obtained by integrating the profile, $N =
\int_{-v}^{+v} N_a(v) dv $. We also computed the  average line centroids and the velocity dispersions 
through the first and second moments of the AOD 
$\bar{v} = \int_{-v}^{+v} v \tau_a(v) dv/\int_{-v}^{+v} \tau_a(v)dv \,\, {\rm km\,s^{-1}}\,, $
and $b = (2 \int_{-v}^{+v} (v -\bar{v})^2 \tau_a(v) dv/\int_{-v}^{+v} \tau_a(v)dv)^{0.5} \,\, {\rm km\,s^{-1}}$,
respectively.
Note that the equivalent widths were measured over the same velocity range $[-v,+v]$ indicated in 
column 8 of Table~\ref{t2}. 

\begin{figure}[tbp]
\epsscale{0.9} 
\plotone{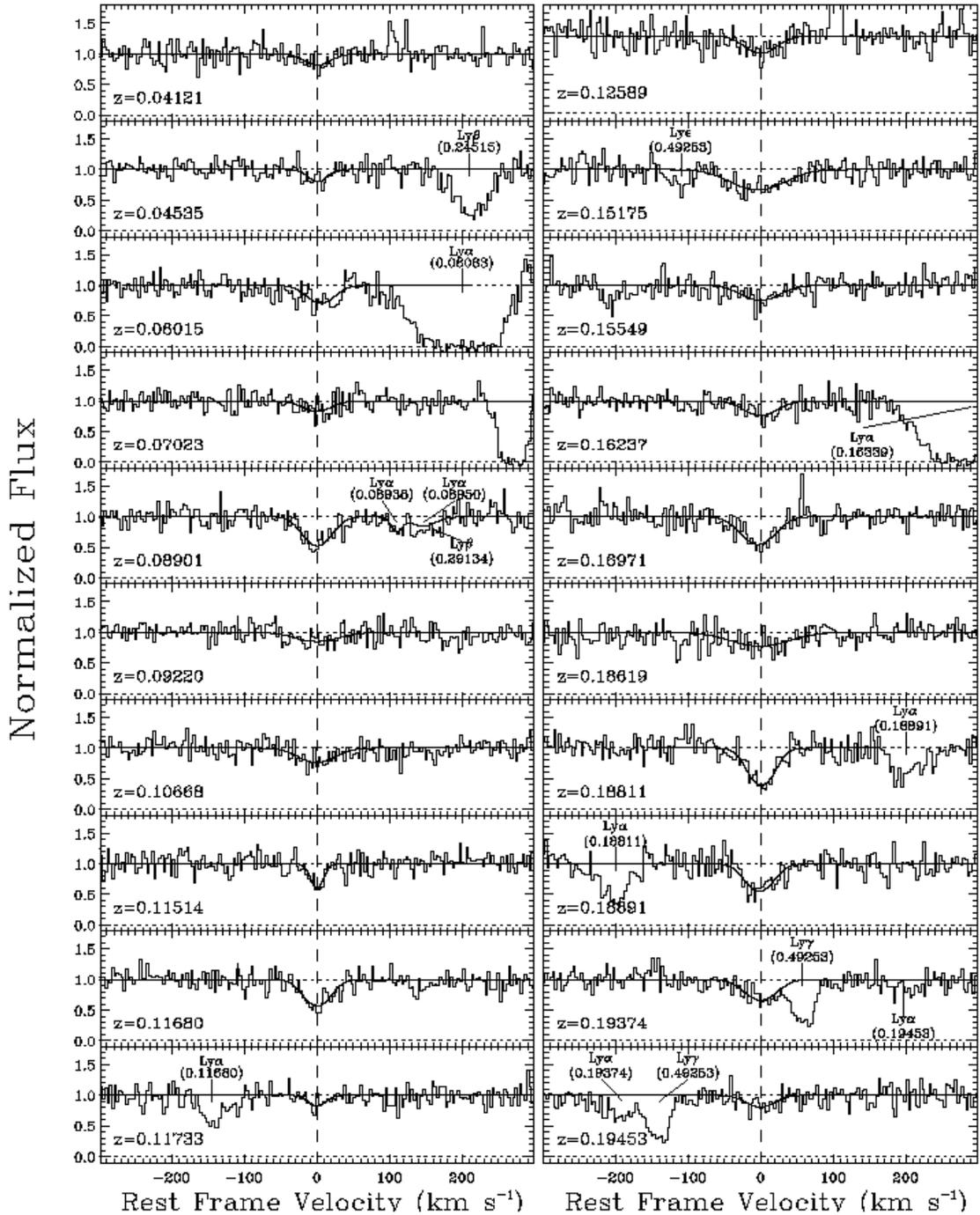}
\caption{\hi\ intervening absorbers detected only in Ly$\alpha$. The continuum normalized fluxes
are plotted against the rest-frame velocity. The redshifts of each absorber are indicated. 
Profile fits with one component to each intervening Ly$\alpha$
are overplotted as solid lines (see text for details). Other lines
present are identified. Note that the systems at $z = 0.08938$ and $z = 0.08950$ 
are shown in the panel for  $z = 0.08901$. Note that the fit for the system
at $z = 0.08950$ was realized by first removing the blend \hi\ $\lambda$1025 at $z=0.29134$.
\label{hi1215only}}
\end{figure}

\begin{figure}[tbp]
\epsscale{1} 
\plotone{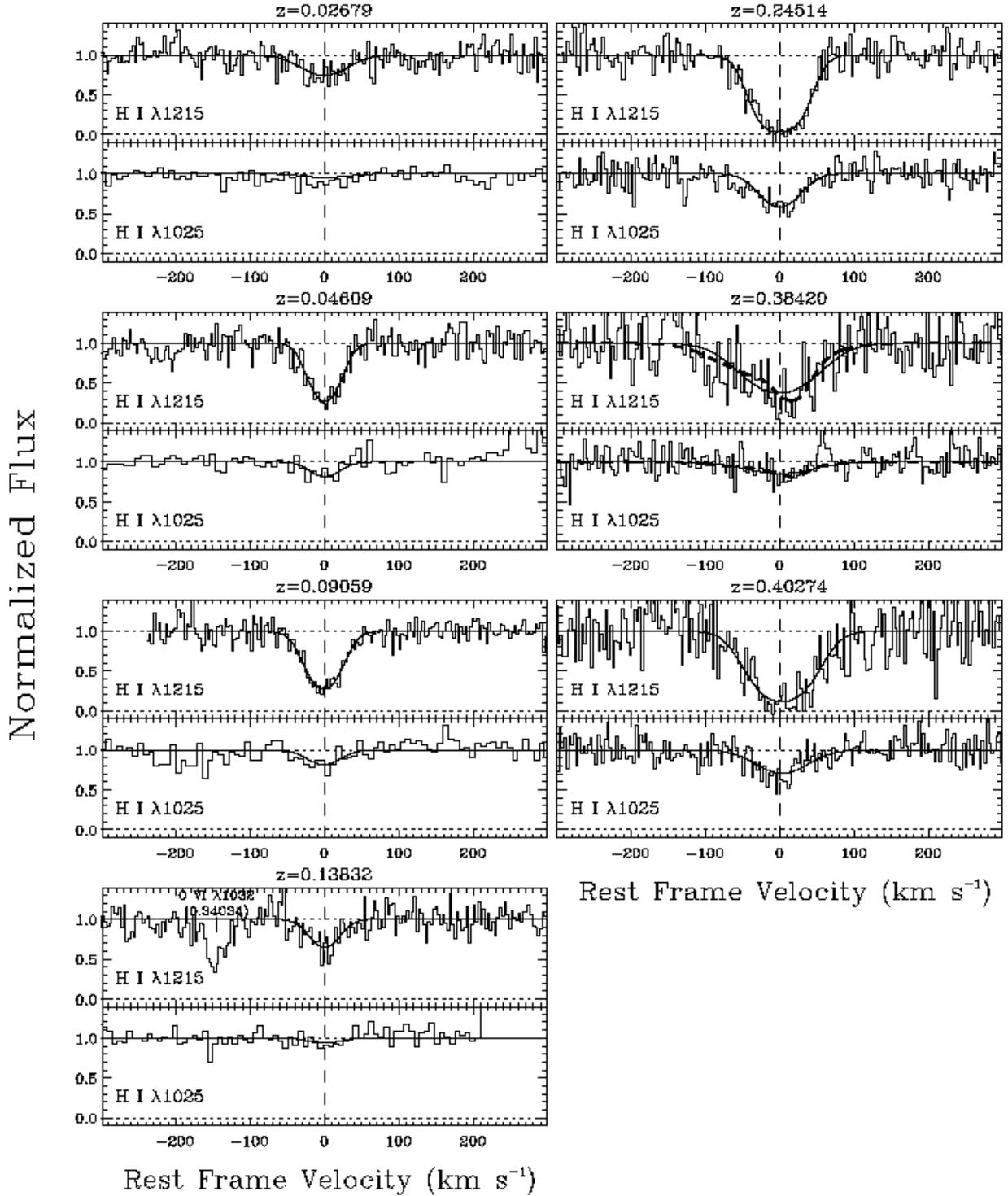}
\caption{\hi\ intervening absorbers detected in Ly$\alpha$ and $\beta$. The continuum normalized fluxes
are plotted against the rest-frame velocity. The redshifts of each absorber are indicated. 
Profile fits with one component to each intervening absorber
are overplotted as solid lines. For the system at $z=0.38420$ a 
two component fit is shown in the dashed line (see note in Table~\ref{t2} for more details). Other lines
present are identified. 
\label{hilyab}}
\end{figure}

\begin{figure}[tbp]
\epsscale{1} 
\plotone{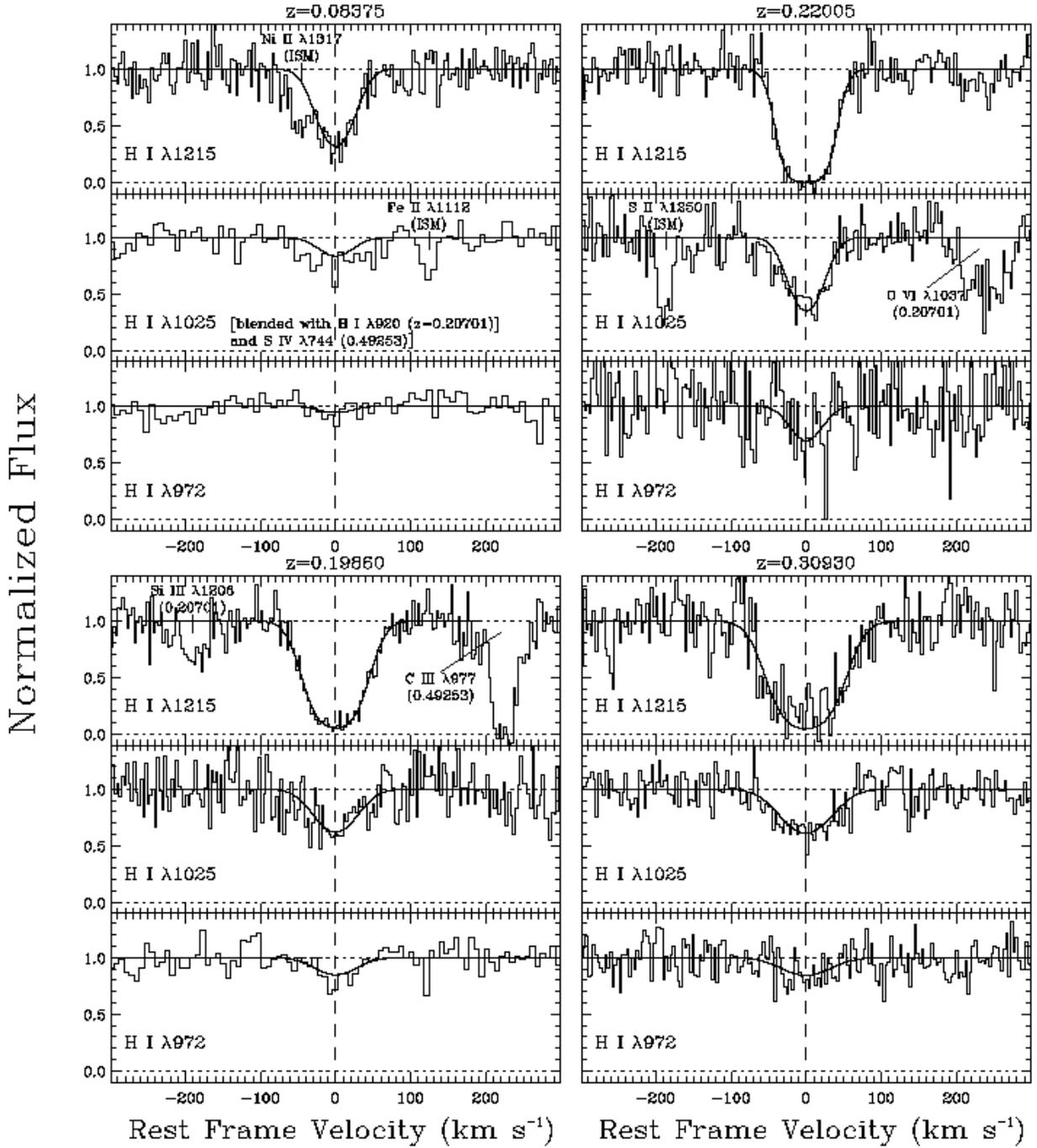}
\caption{\hi\ intervening absorbers detected in Ly$\alpha$, $\beta$, and $\gamma$. The continuum normalized fluxes
are plotted against the rest-frame velocity. The redshifts of each absorber are indicated. 
Profile fits with one component to each intervening absorber
are overplotted as solid lines. Other lines
present are identified. 
\label{hi3}}
\end{figure}

\begin{figure}[tbp]
\epsscale{0.5} 
\plotone{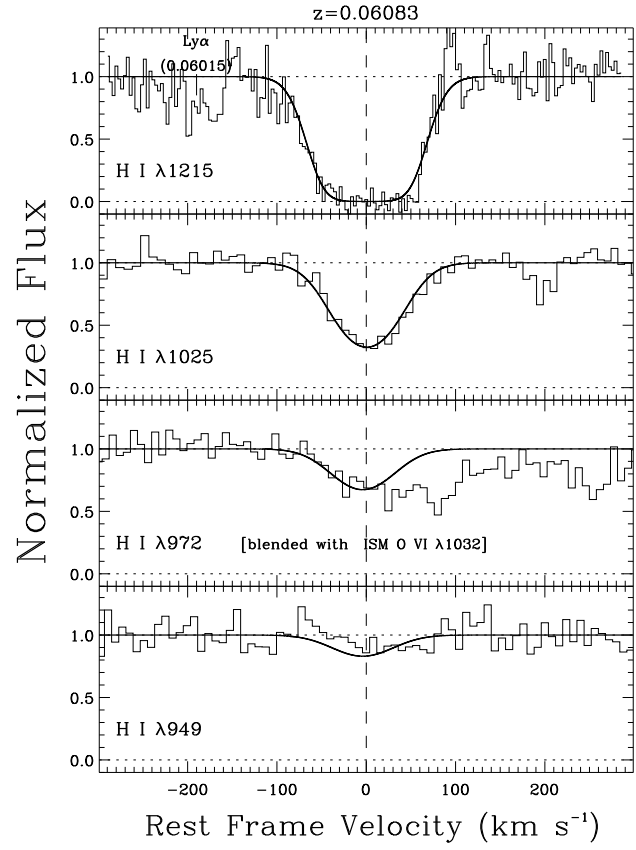}
\caption{\hi\ intervening absorber at $z=0.06083$ detected in Ly$\alpha$, $\beta$, $\gamma$, and 
possibly $\delta$. The continuum normalized fluxes
are plotted against the rest-frame velocity. 
\label{specred060}}
\end{figure}

\begin{figure}[tbp]
\epsscale{0.5} 
\plotone{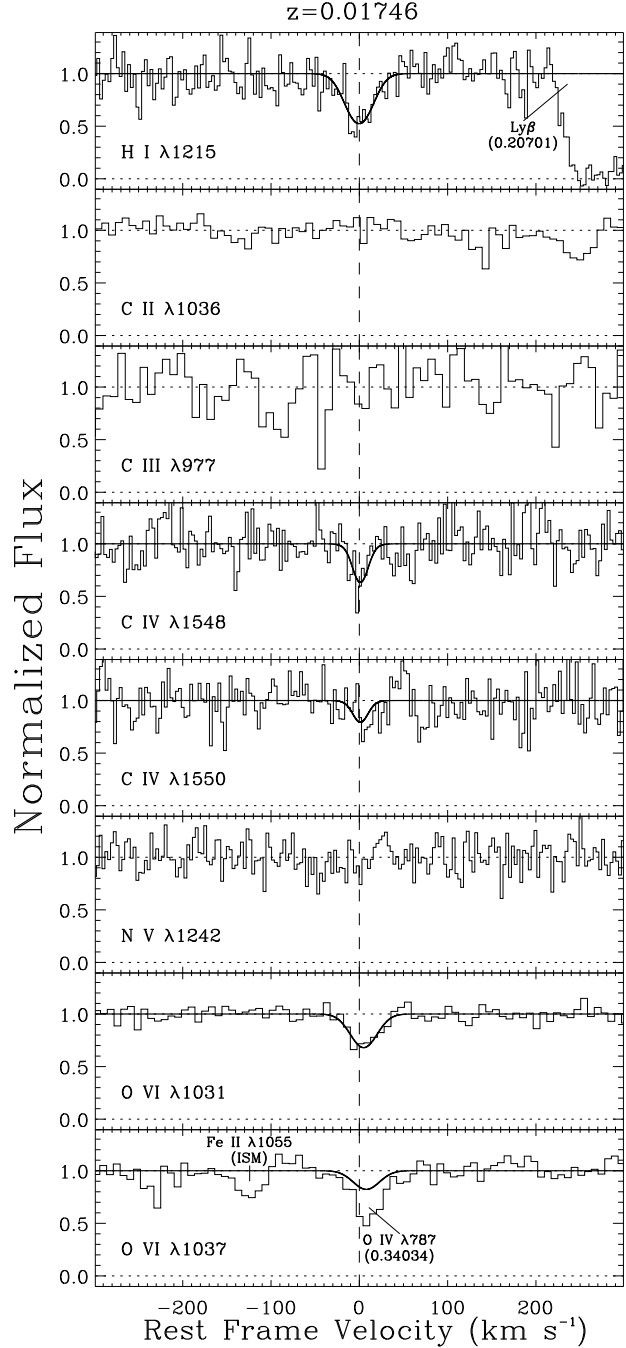}
\caption{Metal-line absorber at $z = 0.01746$.  Profile fits with one component 
are overplotted as solid lines. Other lines
present are identified. \ovi\ $\lambda$1031 is contaminated by H$_2$ LR(1) 4-0 1049.960 \AA,
in the profile that is shown this contaminating H$_2$ line was removed (see note in Table~\ref{t2}).
\label{red0175}}			
\end{figure}

We also fitted the absorption lines with the Voigt component software of \citet{fitzpatrick97}. 
In the {\fuse}\ band, we assume a Gaussian instrumental spread function with a ${\rm FWHM} = 20$ \km,
while in the STIS band, the STIS instrumental spread function was adopted \citep{proffitt02}. Note
that to constrain further the fit to the \ion{H}{1} absorption, we systematically (except as otherwise
stated in Table~\ref{t2}) use non-detected low-order Lyman-series lines (all the lines used in the fit
are listed in Table~\ref{t2} in the line corresponding to the fit result). For this reason, we
favored the $b$-values and column densities derived from profile fitting when realized. We note, 
however, that the  results obtained from the moments of the optical depth and profile fitting methods are  
in agreement to within 1$\sigma$ for most cases.  We show the normalized spectra (with the fit to the absorption lines 
when realized) against the rest-frame velocity of the \hi\ absorbers in 
Figs.~\ref{hi1215only} (systems detected only in Ly$\alpha$), \ref{hilyab}
(systems detected in Ly$\alpha$ and  Ly$\beta$), \ref{hi3} (systems detected in Ly$\alpha$, Ly$\beta$, Ly$\gamma$), 
\ref{specred060} (system detected in Ly$\alpha$,  Ly$\beta$ and  Ly$\gamma$, and Ly$\delta$), and the metal systems in 
Figs.~\ref{red0175}, \ref{red163}, \ref{red340}, \ref{red355}, and \ref{red426}. See also 
Figs.~2 and 4 in \citet{savage05} for the metal system at $z=0.20701$. 
We find several broad \hi\ absorption profiles  with $b> 50$ \km. For most of these systems a 
2  component fit does not improve significantly the reduced-$\chi^2$.   The system at $z=0.38420$
appears to have an asymmetric profile in both the Ly$\alpha$ and Ly$\beta$ lines, and for this
system a 2 component fit looks better by eye but not statistically (see dotted lines in Fig.~\ref{hilyab}). In the note
of Table~\ref{t2} we give the results of the fit for this system: 
a combination of broad ($b = 75$ \km) and narrow ($26$ \km) 
lines seem to adequately fit the \hi\ profiles. Hence, in any case, a broad
component is present. But, only the systems at $z=0.20701$ and 0.27155 
are clearly multiple components blended together. We note that 
several systems are only a separated by less than a few hundreds
\km. We will discuss further these systems in our future paper on the
\hi\ in the low redshift IGM (N. Lehner et al. 2006, in prep.). 

For all the non-detections we list in Table~\ref{t2} the 3$\sigma$ upper limits to the rest-frame
equivalent width and to the column density by assuming the absorption lines lie on the 
linear part of the curve of the growth. We adopted the velocity range  $\Delta v = [v_1,v_2]$ either from other observed metals or from the Lyman lines if no metals were detected (see 
Table~\ref{t2} for more details); except if $\Delta v($\hi$) > 100$ \km, we set $\Delta v= [-50,50]$ \km.

Table~\ref{t3} is a summary table that presents the redshift, the \hi\ column density and Doppler parameter, 
and the column densities of \ciii, \oiii, \oiv, \ovi, and \neviii. The derived \hi\ parameters ($z,N,b$) and the
\ovi\ column densities are from profile fitting. For the other ions, the column density is from AOD 
(except for the columns of the system at $z=0.20701$ that resulted from profile fitting, see Savage et al. 2005). 

\subsection{Possible Misidentification}\label{misid}
While we have done our best to identify properly all the absorption
features in the spectrum of HE\,0226--4110, misidentifications
are possible because we do not have access to the full redshift path
to HE\,0226--4110. The highest redshift at which Ly$\alpha$ 
is detectable is $z=0.423$ (see \S\ref{redpath}). Thus, Ly$\alpha$ 
between 0.423 and 0.495 is not detectable and could produce Ly$\beta$ 
between 1458.6 and 1533.4 \AA, i.e. between redshift $z=0.199$ and $0.261$
(Ly$\alpha$ cannot be Ly$\gamma$ because that would imply $z({\rm Ly}\alpha)\sim 0.54$.)
The following systems are thus potentially affected : $z=0.19860$, 0.20701,
0.22005, 0.22099, 0.23009, 0.23964, 0.24514. However, the 0.19860, 0.20055,
0.20701, 0.22005 and 0.24514 systems have Ly$\beta$ and Ly$\gamma$.
The system at $z=0.20055$ would correspond to Ly$\beta$ at $z=0.42289$. 
At $z=0.42289$, a feature at 1383.822 \AA\ (marked as Ly$\alpha$ at $z=0.13832$
in Fig.~\ref{fig3})  could possibly be identified with Ly$\gamma$, but
the measured \hi\ column densities for Ly$\beta$ and $\gamma$ are discrepant by 0.25 dex. 
Hence, the system at $z=0.20055$ is most likely to be  Ly$\alpha$ too. 
The remaining Ly$\alpha$ systems at $z=0.22099$, 0.23009, 0.23964
could actually be Ly$\beta$ at $z=0.44721$, 0.45799, 0.46931, respectively. 
Those are marked by ``!" in Tables~\ref{t2} and \ref{t3}. 

We note that the $z=0.42289$ \ovi\ system could be Ly$\alpha $ at $z=0.21089$ 
and  0.21756,  or Ly$\beta$ at $z=0.43523$ and 0.44314. However, the derived
physical parameters appear to match very well the atomic
properties of the \ovi\ doublet (see Table~\ref{t2} and Fig.~\ref{red426}).
We also note that several cases where \ovi\ is observed without Ly$\alpha$ or Ly$\beta$
are reported in T. M. Tripp et al. (2006, in prep.):  (i) 
\ovi\ observed without \hi\ is detected in  four cases in associated
systems; (ii) the intervening
system at $z = 0.49510$ toward PKS\,0405--125 has strong \ovi\ with no 
Ly$\beta$,; (iii)  for PKS\,1302--102 at $z = 0.22744$  there is an excellent \ovi\ 
doublet detection and very low ($<2$ sigma) significance Ly$\alpha$ and no Ly$\beta$
detection; (iv) there are several cases in the more complex \hi\ systems where there is clearly 
detected \ovi\ well displaced in velocity from the \hi\ absorption.
Therefore, the $z=0.42289$ system is likely to be an \ovi\ system, but 
we would need a FUV spectrum that covers the wavelength 
up to 1850 \AA\ to have a definitive answer. We will therefore
treat this system in the paper as a tentative \ovi\ system and this  
system is also marked by ``!" in Tables~\ref{t2} and \ref{t3}. 

We finally note that there may be \ovi\ at $z = 0.22005$ and 0.29134. For the system 
at $z = 0.22005$, \ovi\ $\lambda$1037 is identified as Ly$\alpha$ at $z=0.04121$, 
but the \ovi\ $\lambda$1032 appears weaker than expected and is not $3\sigma$. 
Therefore, we do not report these features as \ovi. 
For the system at $z = 0.29134$, \ovi\ would be shifted by $-30$ \km\ 
with respect to \hi. Neither of  the \ovi\ lines are $3\sigma$, 
and therefore we elected not to include them as reliable detections.

\subsection{Unblocked-Redshift Path}\label{redpath}
We will need later (see \S\ref{ofreq}) the unblocked redshift
path for several species under study. 
 The maximum redshift path available for Ly$\alpha$ 
is set by the maximum wavelength available with STIS E140M, 
which is 1729.5 \AA, corresponding to $z_{\rm max} = 0.423$. Note that
the redshift of HE\,0226--4110  is larger, $z_{\rm QSO} = 0.495$. 
The blocked redshift interval arising from interstellar lines, 
other intervening intergalactic absorption lines, and the 
gaps existing in the wavelength coverage is $\Delta z_{\rm B} = 0.022$. 
The unblocked redshift path for \hi\ is $z_{\rm U} = z_{\rm max} - \Delta z_{\rm B} = 0.401$.

For \ovi, we follow a similar method but we note that \ovi\ can arise 
without detection of \hi\ (see \S3.3) since we are not covering the whole 
wavelength range for Ly$\alpha$. We therefore do not restrict the \ovi\ redshift 
path to \hi.  We also restrict the part of spectrum to
where a 3$\sigma$ limit integrated over $[-50,50]$ \km\ is $\la 50$ m\AA. 
Therefore the STIS spectrum between 1182 \AA\ to 1225 \AA\ and above 1565
\AA\ was not used (at $\lambda<1182$ \AA, the {\em FUSE}\ spectrum was used to search 
for \ovi). The unblocked redshift path for 
\ovi\ is $z_{\rm U} = 0.450$. In principle the unblocked redshift path is larger
for the \ciii\ and the EUV lines but because they are much more difficult to 
detect (single line or weaker doublet), we adopt for 
\ciii, \oii, \oiii, \oiv, and \neviii\ the redshift path of 
\ovi\ corrected for any blocked redshift interval arising from interstellar lines and
other intervening intergalactic absorption lines at $\lambda < 1032$ \AA. 
We also consider only wavelengths with $\lambda \ga 924$ \AA\ for those lines because at smaller
wavelengths the spectrum is too confused with the ISM Lyman series absorption 
lines. This would give an unblocked redshift path for \oii\ and \oiii\ of about 0.350 and 
for \oiv\ and \neviii\ of 0.283. 

\section{Physical Conditions in the Metal-Line Absorbers}\label{review}
One of the main issues with the detection of \ovi\ absorbers is given the measurements
of the different species, can we distinguish between photoionization and collisional ionization,
between warm and hot gas? 
This is a fundamental question because to be able to estimate 
$\Omega_{\rm b}($\ovi) in the WHIM, we need to know how much of the observed \ovi\ 
is actually in shock-heated hot gas rather than in cooler photoionized gas. 
For each of the observed absorbers we will first investigate if a photoionization
equilibrium model is a viable option for the source of ionization. We will then 
consider other sources, in particular collisional ionization equilibrium (CIE)
models from \citet{sutherland93}. 

\subsection{Photoionization}
In the IGM, the EUV ionizing radiation field can be energetic
enough to produce high ions in a very low density gas with a long path length.
To evaluate whether or not photoionization can explain the observed properties 
of the \ovi\ absorbers, we have run the photoionization equilibrium code CLOUDY \citep{ferland98} 
with the standard assumptions, in particular that  there has been enough time 
for thermal and ionization equilibrium to prevail. 
We model the column densities of the different ions through a slab illuminated
by the \citet{haardt96} UV background  ionizing radiation field
from QSOs and AGNs appropriate for the redshift of a given system.   
The models assume solar relative heavy element abundances from 
\citet{grevesse98}, but with the updates from \citet{holweger01} for
N and Si, from \citet{allende02} for C, and from \citet{asplund04} for 
O and Ne (see Table~5 in Savage et al. 2005, and see also \S\ref{neviiidis} for Ne). 
With these assumptions, we varied the metallicity and the 
ionization parameter ($U = n_\gamma/n_{\rm H} =$\,H ionizing photon density/total
hydrogen number density [neutral\,+\,ionized]) to search for models that 
are consistent with the constraints set by the column densities of the various
species and the temperature given by the broadening of an absorption line: 
\begin{equation}
T = A(b/0.129)^2
\end{equation}
(where $A$ is atomic weight of a given chemical element). 
The temperature is only an upper limit because mechanisms other than thermal Doppler broadening
could play a role in the broadening of the line.

{\em The System at $z = 0.01746$:}

The absorber system at $z = 0.01748$ (see Fig.~\ref{red0175}) has the lowest redshift of all 
absorbers in our HE\,0226--4110 data. It is detected in Ly$\alpha$, \civ\
($\sim 3\sigma$), and \ovi\ $\lambda$1031 ($8.9\sigma$, \ovi\ $\lambda$1037 
is blended with \oiv\ at $z= 0.34034$). The profile fit to \hi\ 
implies $T < 2.4 \times 10^4$ K. The kinematics appear to be simple 
with all the different species detected at the same velocity within 
$1\sigma$.  These species may therefore arise in the same ionized gas.  To 
investigate this possibility we ran CLOUDY models with $\log N($\hi$)=13.22$.
We estimated that a CLOUDY model with a solar
abundance could reproduce the observed measurements of \hi, \civ\ and \ovi\ and 
the limits on \ciii\ and \nv\ (see Fig.~\ref{cred0175}). The physical 
parameters are tightly constrained by the  \ovi\ column density, with the  
ionization parameter $\log U \simeq -1.1$, the total H density 
$n_{\rm H} \simeq 5.6 \times 10^{-6} $ cm$^{-3}$.
The corresponding cloud thickness is about 17 kpc and the total hydrogen 
column density $3.2 \times 10^{17} $ cm$^{-2}$. The gas temperature is $1.9 \times 10^4$ K, 
similar to the temperature implied by the broadening of the narrow \hi\ line. 
These properties are  reasonable for a nearby IGM absorber: photoionization
is a likely source of ionization for the system at $z = 0.01746$

{\em The System at $z = 0.20701$:}

This system has the highest total \hi\
column density. It is detected in several 
\hi\ Lyman series lines and in many lines of 
heavier elements in a variety of ionization stages, 
with in particular the detection of \neviii. This 
is the most complex metal-line system in the spectrum with at least two velocity-components
in \hi\ and one should refer
to \citet{savage05} for detailed analysis and interpretation
of this system. This system has a metallicity of  $-0.5$ dex and
has multiple phases, including photoionized and shock-heated gas phases. 

{\em The System at $z = 0.34034$:}

This system has absorption seen in Ly$\alpha$, Ly$\beta$, \ciii, \oiv, and \ovi\ 
(see Fig.~\ref{red340}). The velocity-centroids of these species agree well
within the $1\sigma$ errors, and therefore a single-phase photoionized model may 
explain the observed column densities of the various observed species.
We ran the CLOUDY models with $\log N({\mbox \hi}) = 13.68$. 
We estimated that a simple photoionized model with a 1/2 solar
abundance could reproduce the observed measurements of \hi, \ciii\ and \ovi\ and 
the limits (see Fig.~\ref{cred340}, we did not plot the limits for clarity). The physical 
parameters are tightly constrained by the  \ovi\ and \ciii\ column densities, with the  
ionization parameter $\log U \simeq -1.0$, the total H density 
$n_{\rm H} \simeq 1.1 \times 10^{-5} $ cm$^{-3}$. At higher or lower
metallicity, the models cannot reproduce uniquely all the column densities
of the various observed ions. 
The corresponding cloud thickness is about 40 kpc and the total hydrogen 
column density $1.4 \times 10^{18} $ cm$^{-2}$.  The profile fit to \hi\ 
implies $T < (6.7\pm^{2.1}_{1.8}) \times 10^4$ K, which is compatible with the gas temperature
$2.5 \times 10^4$ K derived by CLOUDY at $2\sigma$ with no additional non-thermal broadening.

In conclusion,  the observed simple kinematics and
the column densities can be explained by a photoionization model with a 1/2 solar metallicity,
and $\log U \simeq -1.0$. Therefore, photoionization can be a dominant process for this system
too.

{\em The System at $z = 0.35523$:}

This system is the third most complex metal-system toward HE\,0226--4110, 
with absorption seen in Ly$\alpha$, Ly$\beta$, \oiv, and \ovi. The LiF\,2A
channel suggests a nearly 3$\sigma$ feature for \oiii, but a comparison 
of LiF\,2A and LiF\,1B shows it is only a noise feature (see Fig.~\ref{red355}). 
The \oiv\ feature is, however, real since it is present in both channels, LiF\,1A and  LiF\,2B. 
The profiles of Ly$\alpha$ and Ly$\beta$ are noisy and do not suggest a multicomponent structure. 
\hi\ and \oiv\ align very well, but the \ovi\ profile appears to be more complex. 
The deeper part of the \ovi\ $\lambda$1031 trough aligns well with \oiv,
but there is a positive velocity wing, a 
$\sim$3$\sigma$ feature ($W = 17.0 \pm 5.7$ m\AA) at $\sim$+50 \km. 
This feature could be an extra \ovi\ component or  
a weak intervening Ly$\alpha$ absorption.  
The S/N is not high enough to really understand the 
full complexity of the \ovi\ profile. We therefore treated \ovi\ as being
co-existent with \oiv, and in particular we use the \oiv\ profile to define 
the velocity range for integration of the \ovi\ absorption lines. 

We ran a CLOUDY simulation  with $\log N({\mbox \hi}) = 13.60$. 
The total column density of \oiv\ and \ovi\  and the limit on \ciii\ 
can be satisfied simultaneously in the photoionization model for a very narrow 
range of metallicity near $\log Z/Z_\sun = -0.55 $ at the given $\log N({\mbox \hi}) = 13.60$. 
At a higher metallicity ($\log Z/Z_\sun \ge -0.52 $) the \ovi\ and \ciii\  column density models diverge,
and at a lower metallicity, the models do not predict enough \oiv. In Fig~\ref{cred355},
we show the CLOUDY model with  0.28 solar metallicity. The small error on \ovi\ 
allows only  a very narrow range of ionization
parameters at $\log U \simeq -1.0$ ($n_{\rm H} \simeq 1.1\times 10^{-5} $ cm$^{-3}$). 
The corresponding cloud thickness is about 40 kpc and the total hydrogen 
column density $1.5 \times 10^{18} $ cm$^{-2}$. The gas temperature is $2.9 \times 10^4$ K.
The Doppler parameter of \hi\ implies that $T< (4.4\pm^{2.5}_{1.9}) \times 10^4$ K, which
is compatible with $T$ derived in the CLOUDY simulation. 

For the system at $z = 0.35523$,  a photoionization model with 
0.28 solar metallicity and  $\log U \simeq -1.0$ can explain 
the measured \oiv\ and \ovi\ column densities and the limit on $N($\ciii). 

{\em The System at $z = 0.42660$:}

This system  is only detected in both \ovi\ lines
(see Fig.~\ref{red426}) and could be misidentified (see \S\ref{misid}). 
But, although the \ovi\ $\lambda$1031 line is confused with a spike due to hot pixels, 
both the strength and the separation of the absorption lines match the atomic 
parameters of the \ovi\ doublet. No \hi\ is found associated with this system, 
but the wavelength range only allows us to access Ly$\beta$. 
Because \ovi\ $\lambda$1031 is blended with an emission artifact, it is not clear 
if the \ovi\ profile has only one or more components. 
The negative velocity part of the profile of \ovi\ $\lambda$1031 where the
line is not contaminated by the instrumental spike suggests a rather smooth profile. 
The \ovi\ $\lambda$1037 absorption is noisy and too weak to indicate if more than one component
is needed. We therefore fitted the \ovi\ lines with one component and removed the apparent emission feature 
from the fit. We note that if the fit is made using \ovi\ $\lambda$1037
alone, the parameters are consistent with those of the fit to the doublet. The
errors on the $b$-value may be larger than the formal errors presented in Table~\ref{t2}.
The fit to the \ovi\ absorption lines yields $b = 40.4 \pm 5.0$ \km,
implying $T < 1.6 \times 10^6$ K. 

We explored the possibility that this system may be
principally photoionized by the UV background. Since we do not know the amount 
of \hi\ for this absorber and have only \ovi, 
the results remain uncertain. For a wide variety
of inputs ($\log N($\hi$) = 13.55, 13.30, 13.05$, and $\log Z = [-0.6,0]$), 
the observed \ovi\ column density can be reproduced with a reasonable ionization 
parameter of $\log U \sim -0.7$ or smaller and a cloud thickness less than 100 kpc. 
The broadening of the \ovi\ profiles is non-thermal since photoionization
models give a  gas temperature of  $\la 3 \times 10^4$ K. The Hubble flow broadening 
appears also negligible in most cases. We note that none of the other column density limits constrain
the model further.

\subsection{Collisional Ionization}\label{coll}
We showed above that the \ovi\ systems along with the ancillary ions can be modeled by 
photoionization alone, except for the system at $z=0.20701$ 
described by \citet{savage05} for which \ovi\ clearly traces hot gas. 
If photoionization is the dominant source of ionization of these systems, 
the broadening of \ovi\ and the other metal-lines is dominated
by non-thermal broadening or substructure may be present 
since the temperature of the photoionized gas is typically a few 
$10^4$ K. For \hi, however, there is little
room for other broadening mechanisms since $b_{\rm thermal} \approx b_{\rm total}$. 
We note that \ovi\ is detected at $z= 0.01746$ in 
the {\em FUSE}\ spectrum, which has an instrumental broadening of about $b_{\rm inst} \simeq 12.5$ \km. 
For this system, it is not possible to determine whether $b$ is smaller
than its instrumental width. Within the errors, the broadening of the line 
can be reconciled with a broadening from nearly purely thermal motions. 
The other \ovi\ systems lie in the STIS spectrum, but the S/N of those data is 
not good enough to distinguish between single and multiple absorption components.
In particular, we note the complexity of the \ovi\ profile at $z=0.35523$. 

The broadening of the \ovi\ lines (and \civ\ and \oiv\ 
when detected) implies temperatures of a few $\times 10^5$ K if the broadening is 
purely thermal. At these temperatures, collisional ionization can be an important
source of ionization \citep{sutherland93}. 
If CIE applies, these systems must be multi-phase since
the observed narrow \hi\ absorptions cannot arise in hot gas. Therefore, 
there is a large degree of uncertainty in any attempt to derive 
parameters from CIE models since the fraction of \ovi\ or other ions arising in 
photoionized or collisionally ionized gas is unknown, and the kinematics
do not allow a clear separation of gas phases within different ionization origins. With this caveat
and making the strong assumption that the metal-ions are
not produced in photoionized gas, we review now if CIE could match the observed 
column densities of the metal-line systems:

{\em The System at $z = 0.01746$:}

The broadening of \ovi\ implies $T \sim 2.2\times 10^5$ K (with a large uncertainty)
if it is purely thermal. The limit for $T(\mbox{\ovi})$ is very close to the peak temperature
for \ovi\ in collisional ionization equilibrium \citep[$T = 2.8 \times 10^5$ K;][]{sutherland93}. 
The ratios $\log [N(\mbox{\civ}/\mbox{\ovi})] = -0.5$ is 
compatible with the highly ionized gas being in collisional ionization equilibrium 
at $ T \approx 2.0\times 10^5$ K, close to the thermal broadening for \ovi. 
But $\log [N(\mbox{\nv}/\mbox{\ovi})] < -0.1$ implies 
that N must be subsolar because CIE predicts a fraction of 0.2 at $\log T = 5.30$. 
At this temperature $b($\hi$)_{\rm broad} = 58$ \km\ 
for pure thermal broadening; such a broad component could be superposed 
on the narrow \hi\ absorption and hidden in the noise of the spectrum.
To constrain the \hi\ column density of 
the broad component, we fit the Ly$\alpha$ profile
simultaneously with both narrow and broad lines. For the broad component we fix
$b($\hi$)_{\rm broad} = 58$ \km\ and $v($\hi$)_{\rm broad} \equiv v($\ovi).
The parameters $v,b,N$ are free to vary for the narrow component. We find 
a fit with a very similar reduced-$\chi^2$ as that of the one component fit, giving
$\log N($\hi$)_{\rm broad} = 12.71 \pm 0.37$. In Fig.~\ref{red0175b}, we show 
the fit to the Ly$\alpha$ line for  $\log N($\hi$)_{\rm broad} = 12.71$. 
In CIE, at $\log T \approx 5.30$, the logarithmic ratio of \ovi\ fraction to \hi\ 
fraction is $\log [f({\rm O\,VI})/f({\rm H\,I})] = 3.82 $ \citep{sutherland93}
and the solar oxygen abundance is $\log [{\rm O}/{\rm H}]_\sun = -3.34$ \citep{asplund04},
implying [O/H$]\sim 0.4 $ if $\log N($\hi$)_{\rm broad} = 12.71$ and
[O/H$]\sim 0.0 $ if $\log N($\hi$)_{\rm broad} = 13.10$. Yet, the non-detection 
of \nv\ implies a N/O ratio less than 0.4 times solar. A 
low N/O ratio has recently been observed in a  solar metallicity environment
in the low-$z$ IGM toward PHL\,1811 \citep{jenkins05}. Therefore, this system 
could be collisionally ionized. 

{\em The System at $z = 0.34034$:} 

We find $N($\ovi$)\approx N($\oiv). In CIE, 
\oiv\ and \ovi\ have the same ionic fraction at $T\sim 2.6\times 10^5$ K, which
corresponds to the thermal broadening of these lines within $1\sigma$.  
The fact that $\log [N(\mbox{\nv}/\mbox{\ovi})] < -0.5$ is consistent with this temperature. 
A temperature of $\sim 2.6 \times 10^5$ K implies
$b($\hi$)  = 66 $ \km\ if the broadening is purely thermal.
Such a broad component could be hidden in the noise of the spectrum.
The blue part of the Ly$\alpha$ spectrum 
may indicate the presence of a broad wing (see Fig.~\ref{red340}), but the total recovery 
of the flux on  the red part of the Ly$\alpha$ spectrum indicates
that the features within the blue wing are not part of the main \hi\ absorption. 
While we can force a broad Ly$\alpha$ in the observed profile in the same
way that we did for the system at $z = 0.01746$, the data do not warrant it.
CIE may be viable for this system, but higher quality data are needed to check this. 
The photoionization model with the adopted parameter predicts  $\log N($\nv$) = 13.21$,
while CIE predicts $\log N($\nv$) = 12.75$ (assuming a solar abundance). An increase
in the S/N by about a factor 3 would provide a good 3$\sigma$ limit on \nv\ 
and several other species which would allow a discrimination between
these models. 

{\em The System at $z = 0.35523$:}

The same discussion applies for this system as for the $z = 0.34034$ system since the 
ionic ratios are similar. Higher S/N data would provide better constraints on
\nv\ here as well. However, the broadening of the \oiv\ and \ovi\ 
lines are more consistent with $T \sim 5-6\times 10^5$ K. Therefore 
either there is substructure in the profile as the complexity of
the \ovi\ $\lambda$1031 profile may suggest, or non-thermal broadening
is present if CIE applies. 

{\em The System at $z = 0.42660$:}

The fit to the \ovi\ absorption lines yields $b = 40.4 \pm 5.0$ \km,
implying $T < 1.6 \times 10^6$ K. 
In CIE at $T \sim 1.6 \times 10^6$ K,
we should find  $\log N($\neviii$) = 13.83$ if the relative abundance of Ne and O is solar. 
Our 3$\sigma$ limit for \neviii\ suggests a much smaller column density, less than 13.57 dex. 
If the temperature is $\la 5.4 \times 10^5$ K (corresponding to a thermal $b$-value for 
\ovi\ of about 25 \km), the ratio of \neviii\ to \ovi\ would be consistent with the observed 
ratio (this would also fit the ratio limit for \oiv\ to \ovi). 
If the gas is hot, either the profiles may be more complicated, the broadening not solely
thermal, or the abundance of Ne to  O not solar. We note that the AOD $b$-value
of \ovi\ $\lambda$1037 is less than 2$\sigma$ from 25 \km, where $N($\neviii$)/(N$\ovi$)< 1$. 

{\em The System at $z=0.16339$}

\begin{figure}[tbp]
\epsscale{0.5} 
\plotone{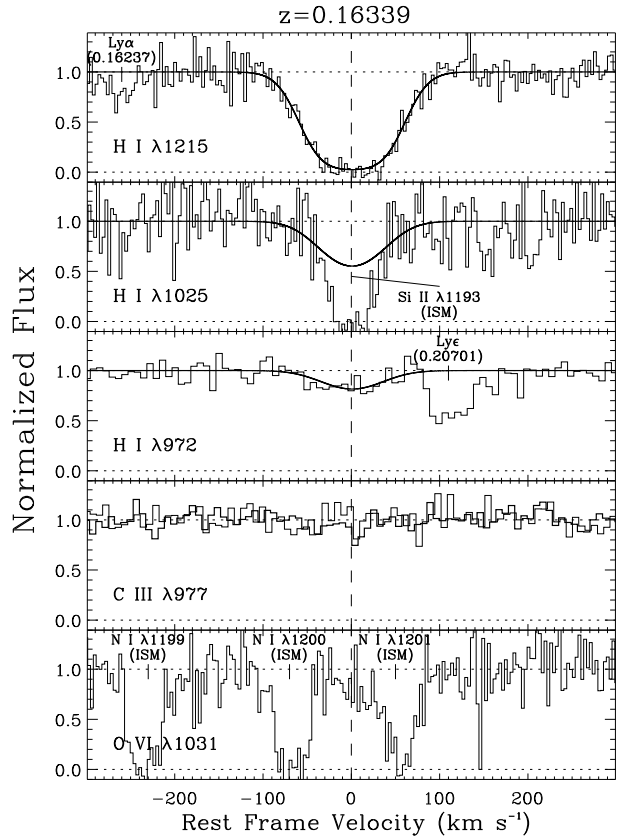}
\caption{Metal-line  absorber at $z = 0.16339$.  Profile fits with one component 
are overplotted as solid lines. Other lines
present are identified. \ciii\ $\lambda$977 
is possibly present. {\em If the 
\hi\ line is thermally broadened and CIE applies}, $Z/Z_\sun \la 0.01$ in
this system. 
\label{red163}}
\end{figure}

\begin{figure}[tbp]
\epsscale{1} 
\plotone{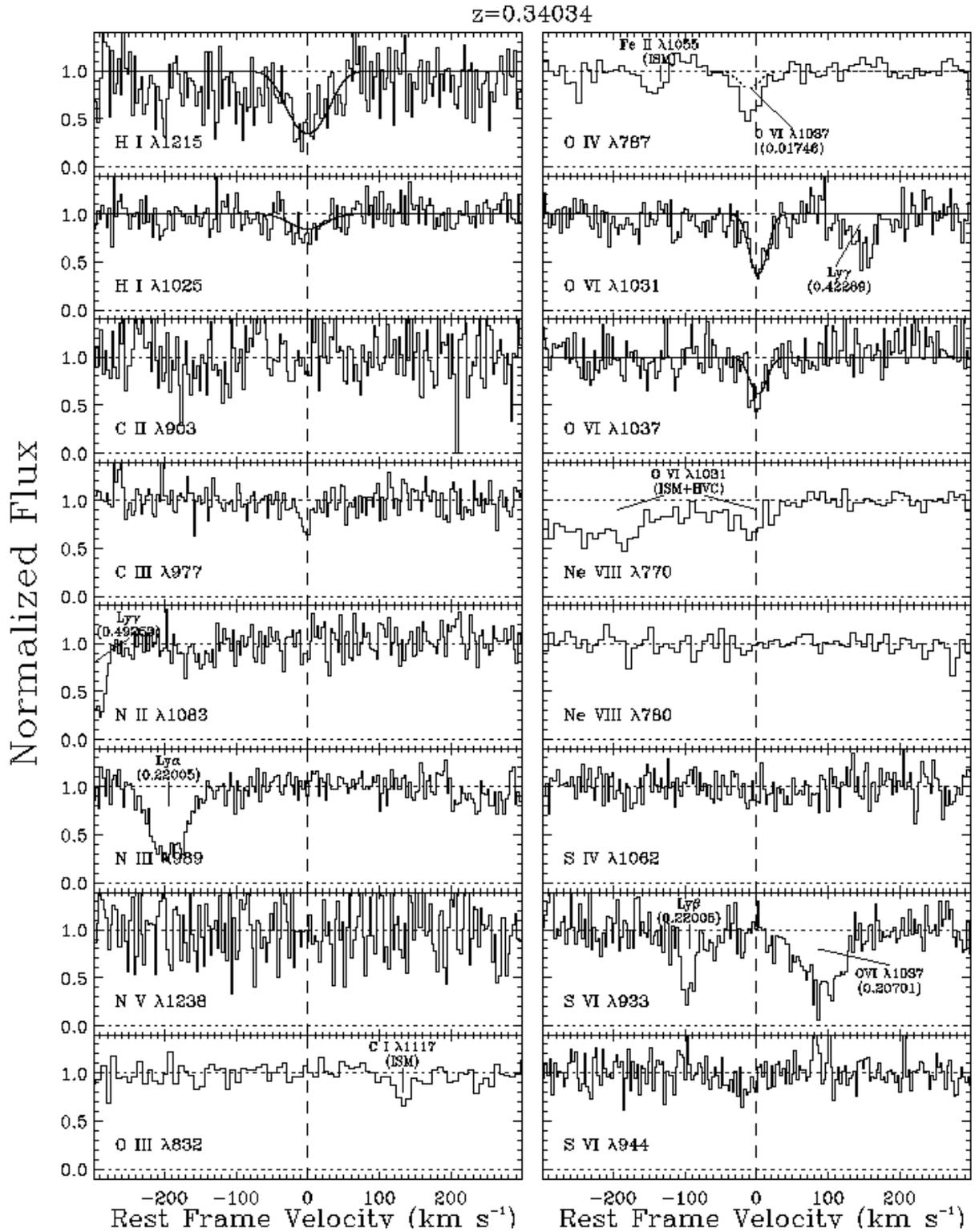}
\caption{Metal-line  absorber at $z = 0.34034$. Profile fits with one component 
are overplotted as solid lines. Other lines
present are identified.  
\label{red340}}
\end{figure}

\begin{figure}[tbp]
\epsscale{1} 
\plotone{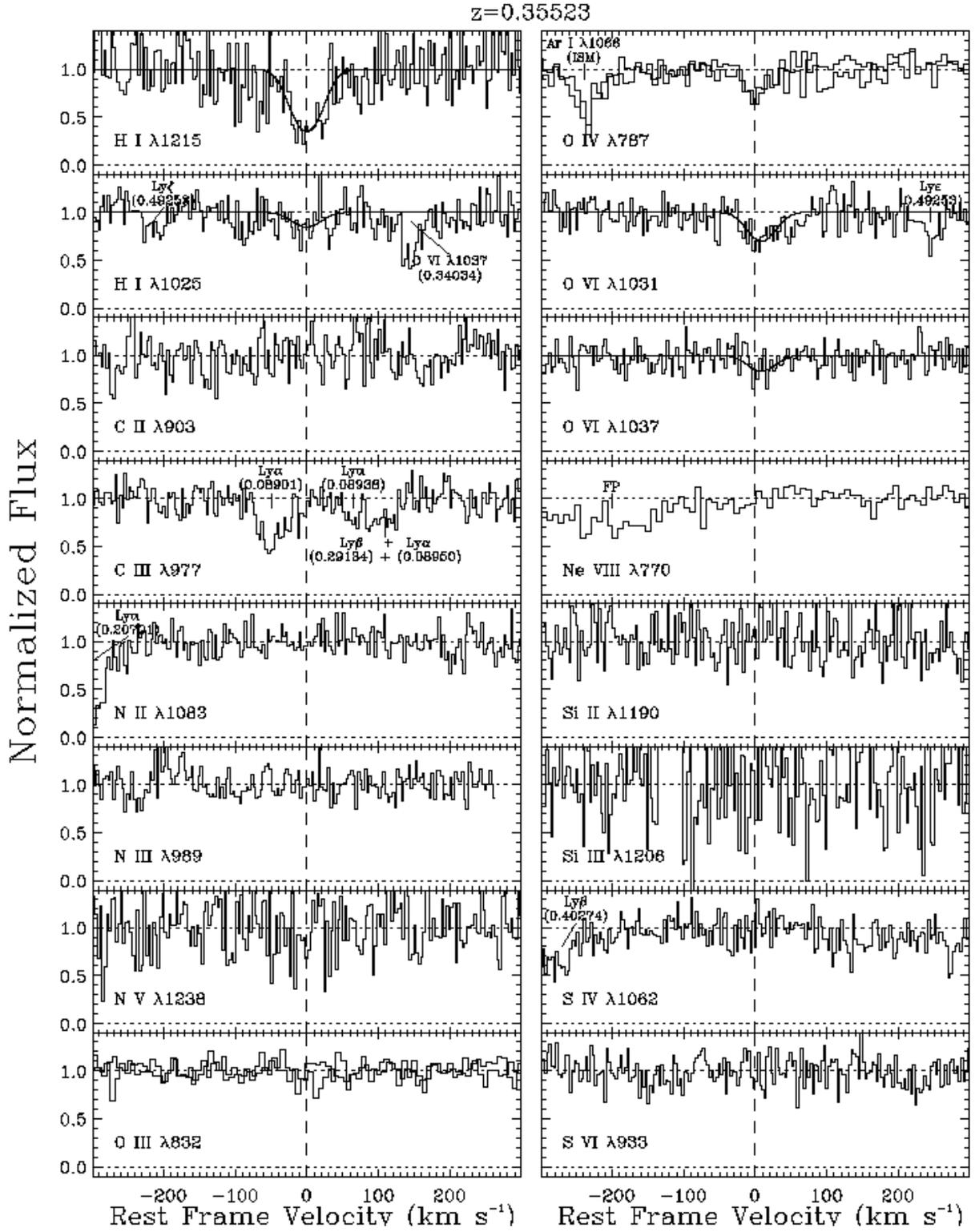}
\caption{Metal-line  absorber at $z = 0.35523$. Profile fits with one component 
are overplotted as solid lines. Other lines
present are identified. The bold spectra in the panels
of \oiii\ and \oiv\ are LiF\,1A and LiF\,2A, respectively. The overplotted
spectra in the panels of \oiii\ and \oiv\ are LiF\,2B and LiF\,1B, respectively.
\label{red355}}
\end{figure}

\begin{figure}[tbp]
\epsscale{0.5} 
\plotone{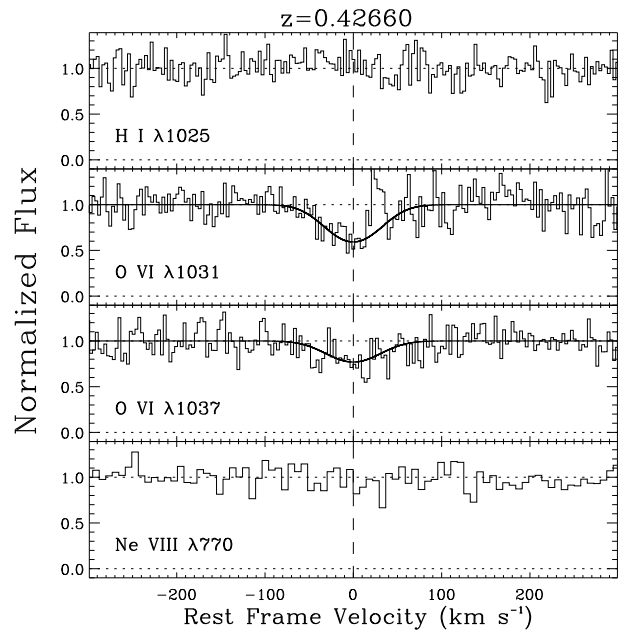}
\caption{Metal-line  absorber at $z = 0.42660$.  Profile fits with one component 
are overplotted as solid lines.
\label{red426}}
\end{figure}

\begin{figure}[tbp]
\epsscale{0.8} 
\plotone{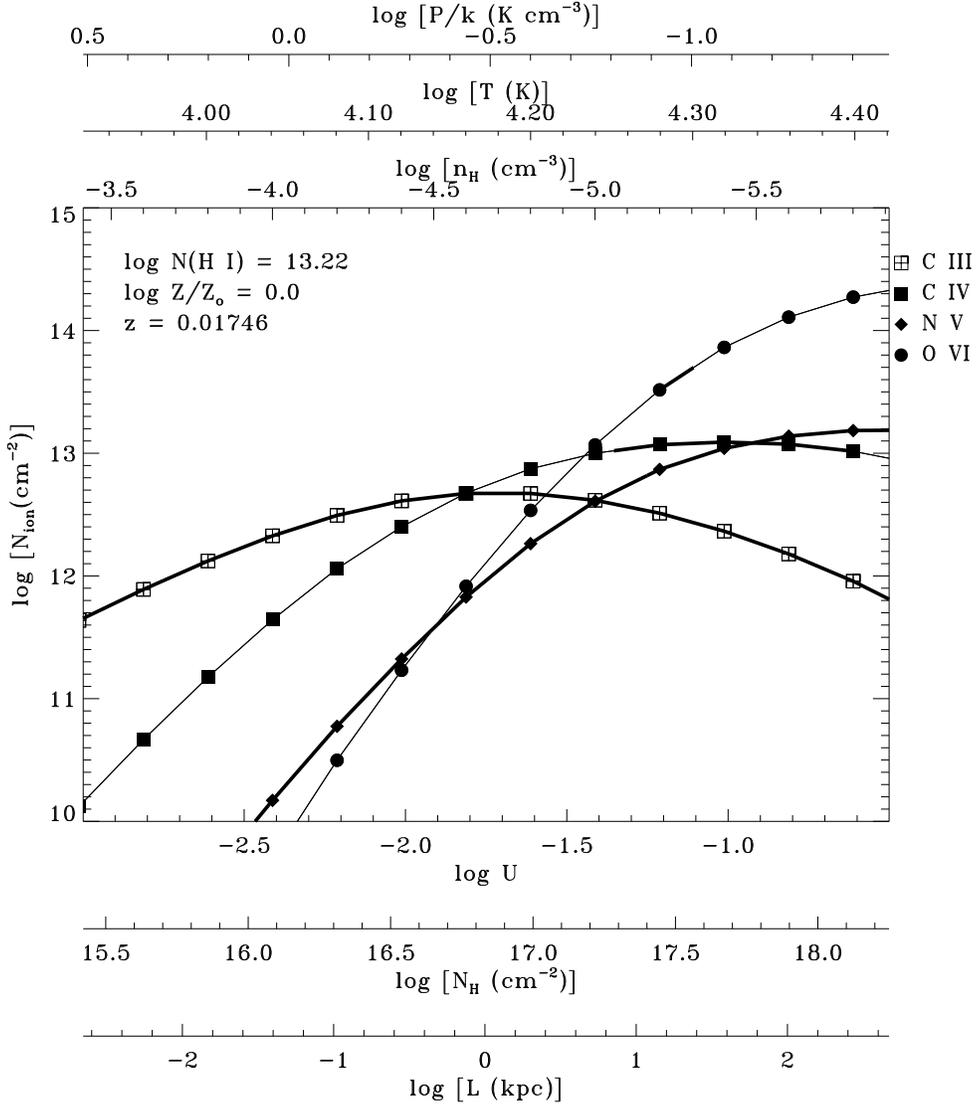}
\caption{Predicted column densities for the photoionization model of the $z=0.01746$ absorber for a 
solar metallicity and $\log N($\hi$) = 13.26$ dex. Pressure,
temperature, density, ionization parameter, total hydrogen column density, and cloud thickness
are plotted along the $x$-axes. The thick solid lines along the thin model curves show the ionization
parameter ranges which are consistent with the observed column
densities (within their 1 sigma uncertainties); the ions corresponding
to each curve are indicated by the key at the right.
 \label{cred0175}}
\end{figure}

\begin{figure}[tbp]
\epsscale{0.8} 
\plotone{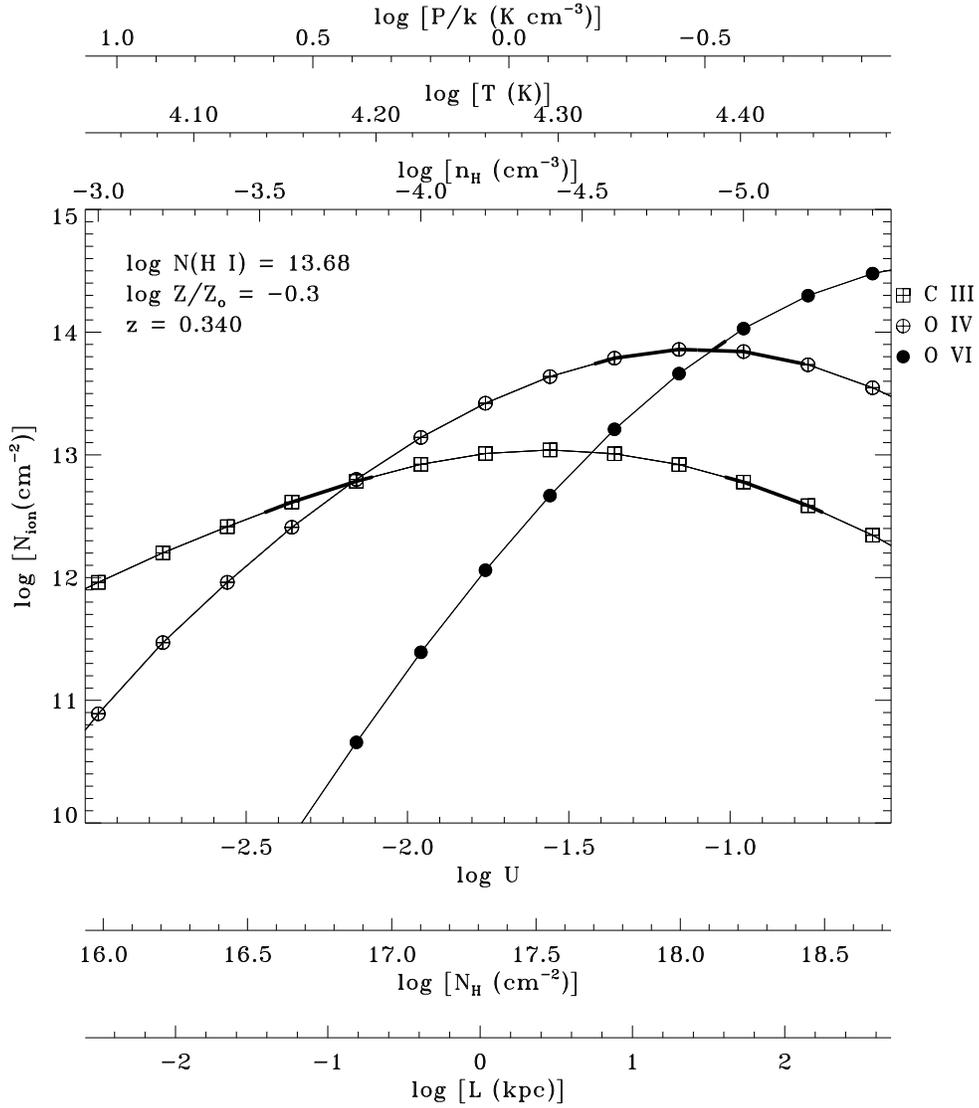}
\caption{Predicted column densities for the photoionization model of the $z=0.34034$ system 
for 1/2 solar metallicity. Axes are defined in Fig.~\ref{cred0175}.
Observed column densities with 1$\sigma$
uncertainties are shown in the thick solid lines along the thin model
curves for each ion.
 \label{cred340}}
\end{figure}

\begin{figure}[tbp]
\epsscale{0.8} 
\plotone{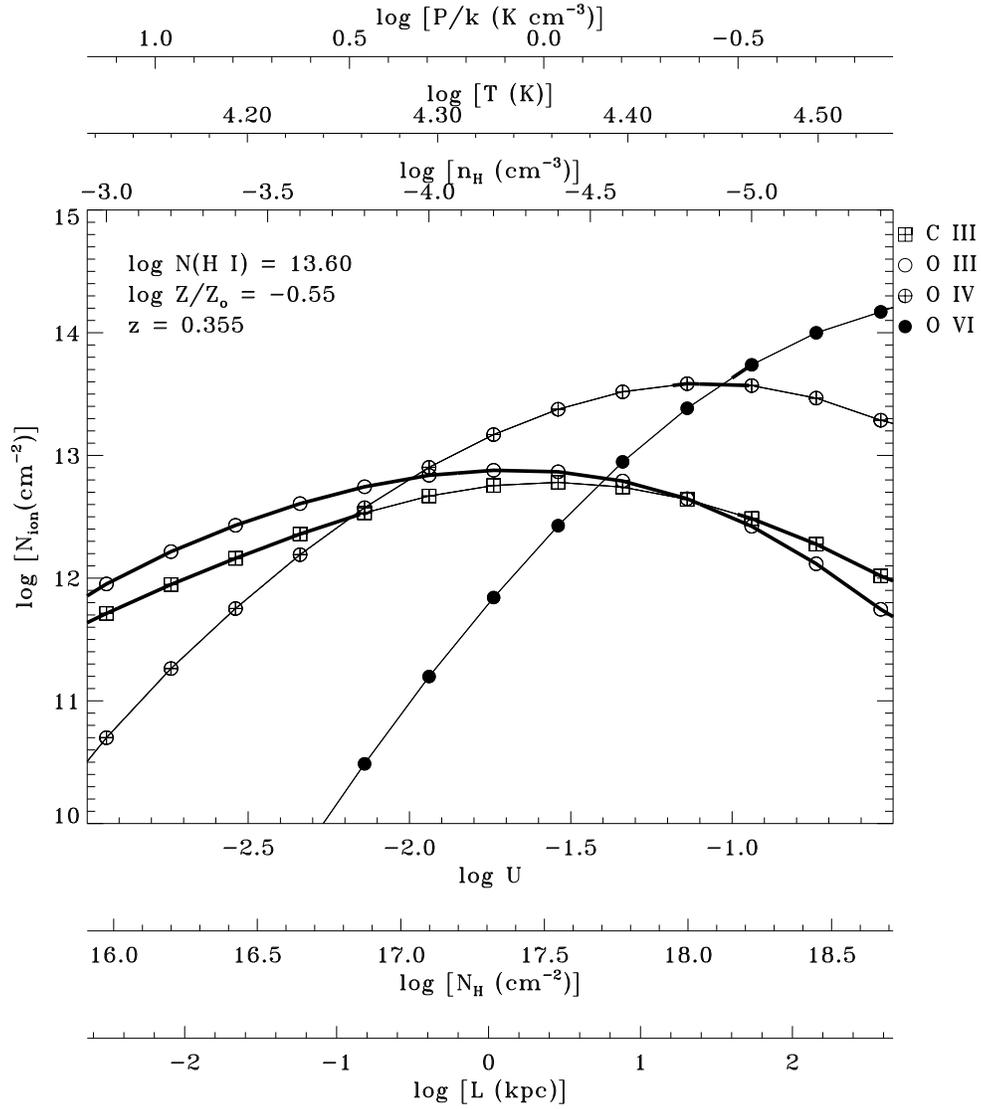}
\caption{Predicted column densities for the photoionization model of the $z=0.35523$ system. 
Axes are defined in Fig.~\ref{cred0175}.
Observed column densities with 1$\sigma$
uncertainties are shown in the thick solid lines along the thin model
curves for each ion.  \label{cred355}}
\end{figure}

\begin{figure}[tbp]
\epsscale{0.7} 
\plotone{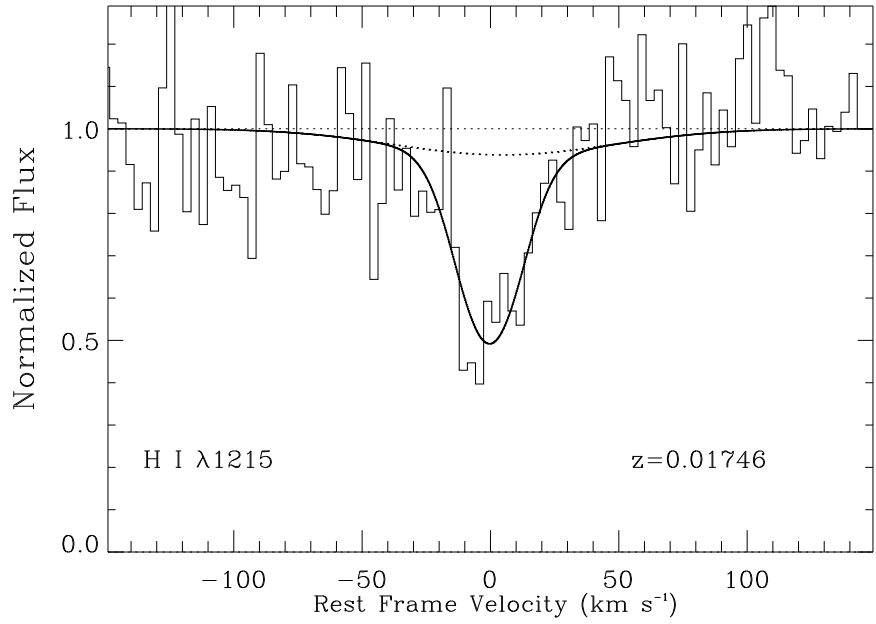}
\caption{Broad component fit to the Ly$\alpha$ line at $z = 0.01746$.
The solid line is a two component 
fit with  $b = 58$ \km\ and $\log N($\hi$) = 12.71$ for the broad component. 
 The dotted line shows the broad component profile only. 
                   \label{red0175b}}
\end{figure}

This absorber has not been discussed yet because no \ovi\ is detected and
there is only a tentative measurement of \ciii.
It has also a broad \hi\ component, potentially tracing hot gas. 
It is seen in Ly$\alpha$ and Ly$\gamma$, with Ly$\beta$
hidden by interstellar \ion{Si}{2} $\lambda$1193  (see Table~\ref{t2} and Fig.~\ref{red163}). 
\oiii\ is confused with H$_2$ (see Table~\ref{t2}). There is a $2.9\sigma$ 
detection of \ciii\ in LiF\,1A.  The data from LiF\,1B have lower
S/N than LiF\,2A and imply a 2.9$\sigma$ upper limit of 12.51 dex for \ciii,
although one pixel is aligned with \ciii\ in LiF\,2A (see Fig~\ref{red163}). 
Within the $1\sigma$ errors, \hi\ and \ciii\ have compatible redshifts. The profile of \hi\ 
is very well fitted with a single Gaussian with $b = 46.3 \pm 1.9$ \km. If the 
broadening is purely thermal, this would imply $T($\hi$) = 1.29 \times 10^5$ K. 
This is the only broad \hi\ system for which a metal ion is (tentatively) detected
along this line of sight.  
Generally, \ovi\ is a more likely ion to associate with  broad \hi\ 
\citep{richter04,sembach04}, but ions in lower ionization stages can
constrain the metallicity of  broad Ly$\alpha$ absorbers as well. We find $[{\rm C/H}] < -2$,
{\em if CIE and pure thermal broadening apply}. 

\subsection{Summary of the Origin of the \ovi\ systems in the Spectrum of HE\,02260-4110}

In summary, the \ovi\ systems at $z = 0.01746$, 0.34034, 0.35523, 0.42660
can be explained by photoionization models. In Table~\ref{t4}, 
we summarize the basic properties of the observed \ovi\ 
systems assuming photoionization and the possible origins of 
these systems.  The broadening of the metal lines
appears to be mostly non-thermal if solely photoionization applies
or there may be unresolved sub-structure buried in the noise of the spectrum. 
Only the system at $z=0.20701$ described by \citet{savage05} appears
to be clearly multiphase with photoionized and collisionally ionized gas. 
The ionic ratios observed in the system at $z = 0.01746$ can
be reconciled with a single CIE model if the relative abundances are non-solar.
For the other systems, CIE is possible with
current constraints. In particular, only  a little non-thermal broadening
would be needed to explain the broadening of the metal lines. Higher
S/N FUV spectra would provide better understanding of the shape
and kinematics of the observed profiles and access to other key
elements such as \nv.

\section{Implications for the Low Redshift IGM}\label{discuss}

\subsection{Discriminating between Photoionization and Collisional Ionization}\label{iondiscuss}
The analysis in \S\ref{review} shows that it is difficult to clearly 
differentiate between photoionization and collisional ionization
for several of the metal-line systems. 
\citet{savage05} showed that the system  at $z = 0.20701$  consists of a mixture of photoionized 
and collisionally ionized gas. This is the only system along this line of sight 
for which \ovi\ cannot be explained by photoionization. We showed that the system
at $z=0.01746$ can be collisionally ionized if N/O is sub-solar, but the observed column densities
can also be explained by a simple photoionization model with relative solar abundance. 
For the \ovi\ systems at $z=0.34034$
and 0.35523 and the tentative \ovi\ system at $z = 0.42660$, photoionization is also the likely 
source of ionization, although collisional ionization cannot be ruled out.

If the photoionization dominates the ionization in a large fraction of the \ovi\ systems, 
estimates of the baryonic density of the WHIM as 
traced by \ovi\ would need to be reduced by a similarly large factor (a factor 6 for 
the HE\,0226--4110 line of sight).
The recent observations in the FUV of the lines of sight toward H\,1821+643 \citep{tripp01}, 
PG\,1116+215 \citep{sembach04}, 
PG\,1259+593 \citep{richter04}, and PKS\,0405--123 \citep{prochaska04} have
similarly shown that the \ovi\ systems are complex, with a mix of photoionized,
and collisionally ionized systems. 

\begin{figure}[tbp]
\epsscale{1} 
\plotone{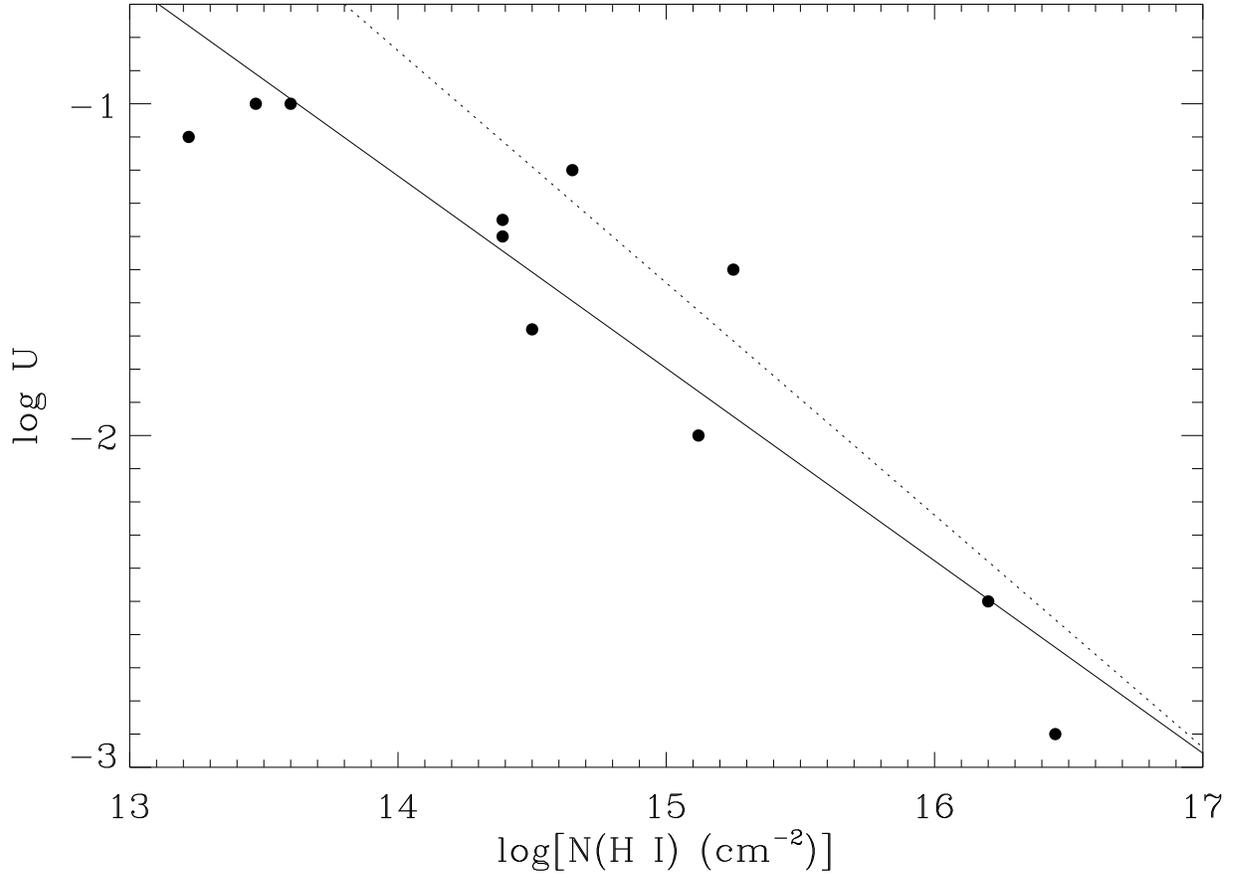}
\caption{The ionization parameter derived from CLOUDY models is plotted against the measured 
\hi\ column density for the systems that could be mostly photoionized.
We combine data from the present work and from the analysis of several other low
redshift QSOs (see \S\ref{iondiscuss} for more detail). 
The solid line is a fit to the data with a slope of $-0.58$, while the cosmological simulation
predicts a slope of $-0.7$ (dotted line).
\label{hirel}}
\end{figure}

\citet{prochaska04}  noted a general decline of the  photoionization parameter $U$
estimated in the CLOUDY model with increasing observed \hi\ column density. 
\citet{dave99} found that $n_{\rm H} \propto N($\hi$)^{0.7} 10^{-0.4 z}$ for the low-$z$ Ly$\alpha$
absorbers. Assuming that the \hi\ ionizing intensity is constant gives
$U \propto N($\hi$)^{-0.7}$ \citep{prochaska04}. To check if this trend holds
with more data and in particular  to check if the \ovi\ systems that could be explained
by both photoionization and CIE deviate from this trend, in 
Fig.~\ref{hirel} we combine 
our results summarized in Table~\ref{t4} for the photoionized \ovi\ systems with results from 
\citet{prochaska04}, \citet{richter04}, \citet{savage02}, and \citet{sembach04}. 
We show in Table~\ref{t4a} the redshift, $\log U$, and  
\hi\ column density.  A linear fit using a least squares fit
to the data gives  $\log U = -0.58 \log (N({\mbox\hi})/10^{14}) -1.23 $ (we did not include
the peculiar \ovi\ system at $z=0.36332$ toward PKS\,0405--123 where the very weak Ly$\alpha$
is offset from the metal-line transitions, see Prochaska et al. 2004; including this system would change
the slope to $-0.55$). For the fit we treated the lower limit of $\log U$ toward PKS\,0405--123 
as an absolute measure, but excluding this limit from the fit would not change the result. The solid curve 
in Fig.~\ref{hirel} shows the fit.
While the slope of $-0.58$  is close to the predicted slope of $-0.7$ from the numerical simulations 
\citep{dave99,dave01}, the simulations appear to provide an upper limit to this relation (see dotted line
in Fig.~\ref{hirel}).  
Note that metal-line systems with no \ovi\ do not seem to follow this correlation. For example the system at 
$z=0.00530$ toward 3C\,273 gives $\log N($\hi$)=15.85$ and $\log U = -3.4$ \citep{tripp02}, 
and the system at $z=0.16610$ gives $\log N($\hi$)=14.62$ and $\log U = -2.6$ \citep{sembach04}: 
both systems are very much below the distribution of the data plotted in Fig.~\ref{hirel}.
Although the correlation 
could be somewhat fortuitous for the \ovi\ systems that can be explained by both
photoionization and CIE origins, it would have to occur for all these systems 
(e.g., 2 systems presented in this paper, and the
system at $z = 0.14232$ toward PG\,0953+415 in Savage et al. 2002). 
This may favor photoionization for the \ovi\ systems that can be fitted by both photoionization
and CIE models. 

\begin{figure}[tbp]
\epsscale{1} 
\plotone{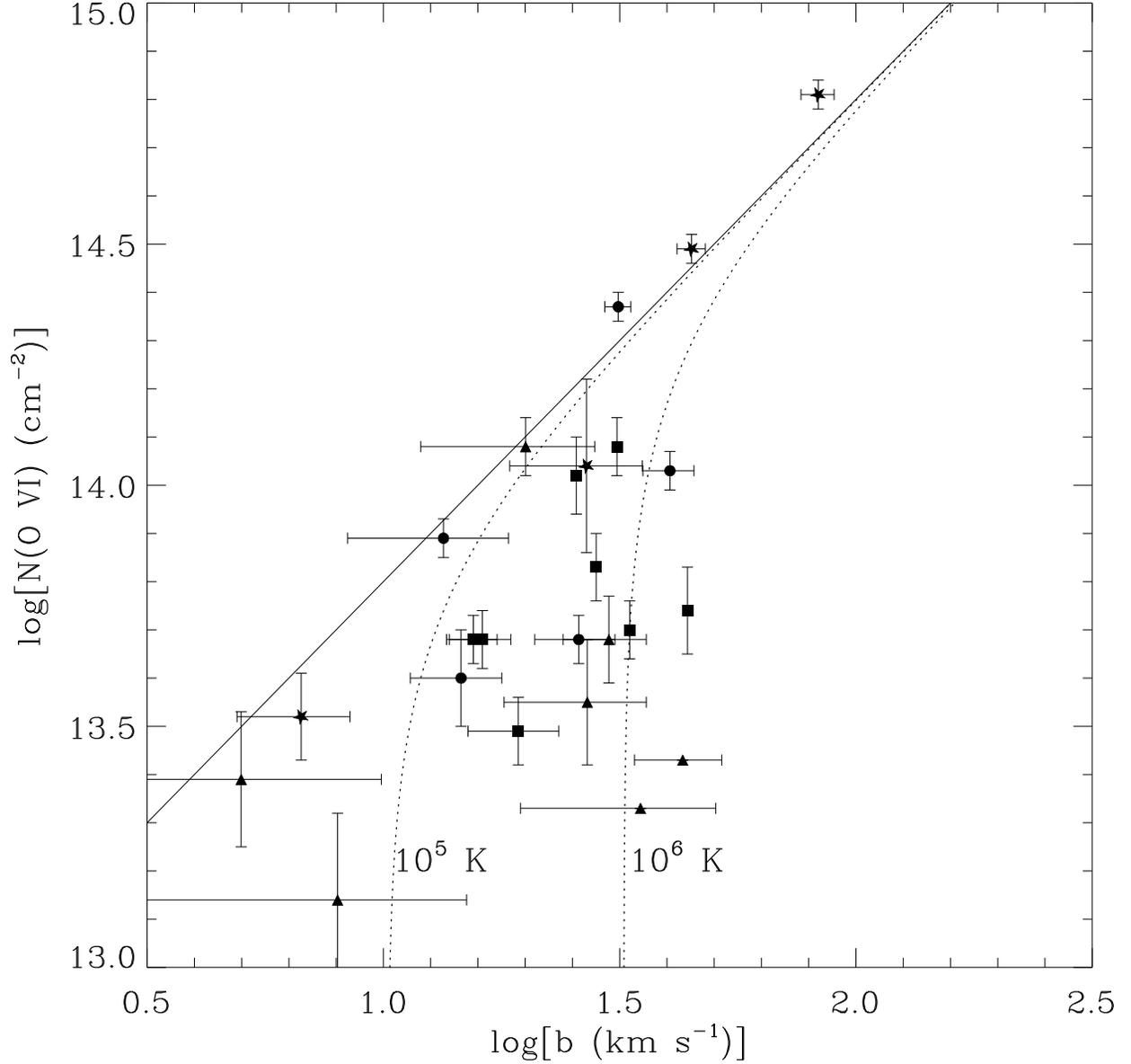}
\caption{Column density vs. the Doppler parameter for the intervening \ovi\ absorption systems 
observed toward HE\,0226--4110 ({\em circles}, this paper), PKS\,0405--123 ({\em stars}, Williger et al. 2005),
PG\,1116+215 ({\em triangles}, Sembach et al. 2004), and PG\,1259+593 ({\em squares}, Richter et al. 2004).
The relation predicted by \citet{heckman02} for radiatively cooling gas is shown in dotted lines for assumed
\ovi\ temperatures $10^5$ K and $10^6$ K.  
The solid line corresponds to the linear regime where identical components have the same central velocity 
\citep[$\Delta v = 0$ \km\ in][]{heckman02}.  
\label{novib}}
\end{figure}

So far, we have mainly considered photoionization and CIE models to explain 
the observed \ovi\ systems. However, if the gas is hot (between $10^5 $ 
and $10^6$ K) and dense enough, it should cool rapidly since this 
is the temperature range over which the cooling of an ionized plasma is
maximal \citep{sutherland93}.  \ovi\ is the dominant coolant in
this temperature range. \citet{heckman02}  investigated the non-equilibrium 
cooling of \ovi\ and computed the relation between 
$N$(\ovi) and $b_{\rm obs}$(\ovi) expected for radiatively cooling gas at 
temperatures of $10^5 $ and $10^6$ K. 
Considering the 4 IGM sight lines that have now been fully analyzed
(HE\,0226--4110, this paper; PKS\,0405--123, Williger et al. 2005;
PG\,1116+215, Sembach et al. 2004; PG\,1259+593, Richter et al. 2004), 
we plot in Fig.~\ref{novib} the column density against the observed width for \ovi\ 
absorption systems seen along these lines of sight. Most of the data
lie between the $b$-range [10,35] \km\ and $\log N$-range [13.5,14.1]
with a large scatter. Considering data outside 
these ranges, there is a general trend of the \ovi\ column 
increasing with increasing \ovi\ width, as observed in the Galaxy, 
Small and Large Magellanic Clouds \citep{savage03}. 
There is no obvious separation in Fig.~\ref{novib} between 
systems that could be photoionized or collisionally ionized. 
In this figure, we also 
reproduce Heckman et al.'s radiative cooling models for $10^5 $ 
and $10^6$ K and the $N/b$ (\ovi) linear regime ($\Delta v =0 $ \km, see Heckman et al.).
Their models can reproduce the observed $b-N$ distribution. 
\citet{heckman02} also computed the column densities 
of several ancillary species. We reproduce the observed and predicted ratios of 
\oiv/\ovi, \neviii/\ovi, and \svi/\ovi\ in Table~\ref{t5}. 
Generally, their models produce too much \neviii. 
For \oiv\ and \svi, the comparison may be more difficult because
photoionization may play a role, but photoionization can also 
produce \ovi\ if the IGM has a very low density. So while these cooling models can reproduce
the distribution of $N$(\ovi) and $b_{\rm obs}$(\ovi), they
generally fail to predict observed ionic ratios, especially
\neviii/\ovi. The pure isobaric model is always ruled out, as it predicts
a \oiv/\ovi\ ratio that is too low.
Note that these models assume solar relative elemental abundances.  

In summary, our results show that
for 1 out of 5 \ovi\ systems toward HE\,0226--4110, 
collisional ionization is the likely origin of the \ovi.
For the 4 other systems, comparison of the ionic ratios and the kinematics
of the ionic profiles does not yield  a single solution for
the origin of the \ovi.  We note that $U$ and $N$(\hi)  of these  systems
correlate and follow the same distribution as pure photoionized \ovi\ systems, 
suggesting that photoionization could be 
the origin of these \ovi\ systems.
To better understand the basic properties  of the IGM in the low redshift
universe, observations with
high S/N data are needed. At high redshift, data with S/N of 100 per 7 \km\ resolution element
have answered many fundamental questions that remained unanswered  
with low quality data. While in the near future FUV spectra will not have
the quality of the current optical data, S/N of $\sim$50 or higher could be achieved 
on bright QSOs with the Cosmic Origin Spectrograph (COS) when/if it is installed on the
{\em HST}.

\subsection{Frequency of Occurrence Of Oxygen in Different Ionization Stages}\label{ofreq}

Oxygen is certainly the best metal element for the study of the 
physical properties of the IGM because it is the most abundant
and because it has a full range of ionization 
stages, from \oi\ to \oviii, accessible to existing spectrographs.  While we are far from being able
to combine X-ray (\ovii, \oviii) lines with FUV lines because of 
sensitivity and spectral resolution issues, searching for \oi\ to 
\ovi\ is possible by combining EUV and FUV lines available in 
the {\fuse}\ and STIS wavelength ranges. In Table~\ref{t6},
we summarize the transitions of oxygen observable 
in the EUV--FUV  and the  redshift at which these transitions
can be observed. Toward HE\,0226--4110, we unfortunately miss 
\ov\  $\lambda$629 (because the redshift path
of HE\,0226--4110 is not large enough (\ov\ is detected in the associated
system, see Fig.~\ref{fig2}). The close match in the oscillator strengths
among these ions also allow direct comparisons of lines with similar column 
density sensitivities (see Table~\ref{t6}). 

For the Ly$\alpha$ systems detected toward HE\,0226--4110 it is almost always
possible to search for the associated metal lines, as blending is rarely a
problem. The only system where we cannot study \ovi\ is the one at $z = 0.09059$.
For all the other systems, we are able to measure
the  \ovi\  column density or estimate a 3$\sigma$ limit. For 5 absorbers, 
($z= 0.06083,0.10667,0.16237,0.16971,0.19861$) \ovi\ $\lambda$1031 is blended with 
other IGM or ISM absorptions, so the limits were estimated with \ovi\ $\lambda$1037.  
For \oiii\ and \oiv\, there are only 8 and 2 systems, respectively, 
for which we cannot make a measurement because of blending (see Table~\ref{t3}).  We do not report
estimates for \oi\ and \oii\ because these lines are rarely detected in 
the Ly$\alpha$ forest (they are sometimes detected in the high \hi\ column systems
with $\log N($\hi$)>16.1$ , see Prochaska et al. 2004, Sembach et al. 2004, Tripp et al. 2005). 
Since  no unidentified features with 
$W \ge 3\sigma$ lie at the wavelengths where \oi\ and \oii\ associated with \hi\ 
are expected,  we can confidently say that there are no intervening systems with \oi\ and \oii\ lines in the spectrum
of HE\,0226--4110.

A useful statistical quantity to use for constraining 
the ionization of the IGM is the number of intervening systems per unit redshift.
For \ovi, five intervening systems toward HE\,0226--4110 are detected in either one ($z =  0.01748$)
or both \ovi\ lines ($z =  0.20701, 0.34034, 0.35523, 0.42660$). These 5 systems have 
$W(1031) > 48$ m\AA.  The number of 
intervening \ovi\ systems with $W_\lambda \ga 50$ m\AA\ per unit redshift
is $d{\mathcal N}({\mbox \ovi})/dz = 11$ for an unblocked redshift path of 
0.450 (see \S\ref{redpath}). This number would be 9 if the tentative $z=0.42660$ system
is not included. 

There is one definitive detection of \oiii\ 
and there are 3 detections of \oiv\ in the spectrum of HE\,0226--4110. Using
the redshift path defined in \S\ref{redpath}, $d{\mathcal N}({\mbox \oiii})/dz = 3$ and
$d{\mathcal N}({\mbox \oiv})/dz = 11$. If we adopt the unblocked
redshift path of \hi\ instead of \ovi, these numbers do not change significantly.
While this sample still suffers from small number statistics, it suggests 
$ d{\mathcal N}({\mbox \oiii})/dz < d{\mathcal N}({\mbox \oiv})/dz \approx d{\mathcal N}({\mbox \ovi})/dz $
in the IGM along the HE\,0226--4110 sight line. 
Since the values of $\lambda f$ of the oxygen lines are similar, the column 
density limits for these different ions are similar. 
The ionization potentials of \oiii\ and 
\ciii\ are similar and since there are more systems for which we can search for \ciii,
we also report the measurement of the strong \ciii\ line ($\log \lambda f = 2.85$) in Table~\ref{t3}.
Only for the systems at $z = 0.06083,0.08901,0.24514,0.42660$ were we not
able to estimate the amount of \ciii. We find
$d{\mathcal N}({\mbox \ciii})/dz = 4-6$. \citet{richter04} reported
4 definitive \ovi\ systems and detection of two \oiii\ and \oiv\ systems (one of them
not associated with \ovi) toward PG\,1259+593. Adopting the same procedure as above, toward
PG\,1259+593, we find $ [d{\mathcal N({\mbox \oiii})}/dz = 8] < [d{\mathcal N}({\mbox \oiv})/dz = 10] \approx 
[d{\mathcal N}({\mbox \ovi})/dz = 11] $. There are 5 reported \ciii\ systems against 6 
\ovi\ systems along the sight line toward 
PKS\,0405--123 \citep{prochaska04}, implying $ d{\mathcal N}({\mbox \ciii})/dz \la d{\mathcal N}({\mbox \ovi})/dz  $.
Along the path to PKS\,0405--123, there are 1 \oiii\ system and 4 detected \oiv\ systems that imply
$ [d{\mathcal N({\mbox \oiii})}/dz = 4] < [d{\mathcal N}({\mbox \oiv})/dz = 17] \approx 
[d{\mathcal N}({\mbox \ovi})/dz = 16] $. Combining these sight lines, we find: 
$ [d{\mathcal N({\mbox \oiii})}/dz \approx 5] < [d{\mathcal N}({\mbox \oiv})/dz \approx 13] \approx 
[d{\mathcal N}({\mbox \ovi})/dz \approx 13] $. 

The general pattern that emerges is that there is a slightly larger number of \oiv\ and \ovi\ 
systems per unit redshift compared to \oiii, about the 
same number of \oiv\ and \ovi\ systems per unit redshift, but a much larger number
of \oiii\ systems per unit redshift  than for \oi\ and \oii. The 
low redshift IGM is more highly ionized than 
weakly ionized. This is consistent with the picture of the Ly$\alpha$ forest 
consisting mainly of very low density photoionized gas and hot gas. We also 
note that when \oiii, \oiv, and \ovi\ are detected simultaneously the column densities
of these ions generally do not differ by more than a factor 2--3.  

\subsection{Observations of \neviii}\label{neviiidis}

Li-like \neviii\ provides a powerful diagnostic of hot gas because in CIE 
it peaks at $7\times 10^5$ K, Ne has a relatively high cosmic
abundance ([Ne/H$]_\sun= -4.16$) and the \neviii\ lines have relatively high $f$-values 
($\log f\lambda = 1.90, 1.60 $). In CIE at $T \sim (0.6-1.3 )\times 10^6$ K, $N($\neviii$)/N($\ovi$) > 1$.
For $T \sim (0.6-1.0)\times 10^6$ K, $N($\neviii$)\sim (2-3)\times N($\ovi$) $. \citet{savage05}
discussed the system at $z=0.20701$ in the spectrum of HE\,0226--4110 
and they reported the first detection of an intervening \neviii\ system
in the IGM. The $z=0.20701$ 
\ovi/\neviii\ system arises in hot gas at $T \sim 5.4 \times 10^5$ K. 
With little contamination from H$_2$ lines and its high redshift path,
HE\,0226--4110 currently provides the best line of sight to search for \neviii.
We found, however, only one detection of \neviii\ among 
the 32 \hi\ systems (excluding the associated system) at redshifts 
where \neviii\ could be measured.  We also searched for 
pairs of absorption features with the appropriate separation of the 
\neviii\ doublet, not associated with a Ly$\alpha$ feature, and found none. 
Note that none of the \hi\ systems observed are broad enough to produce a
significant fraction of \neviii, since the broadest \hi\ implies $T< 2.8\times 10^5$ K.

\begin{figure}[tbp]
\epsscale{1} 
\plotone{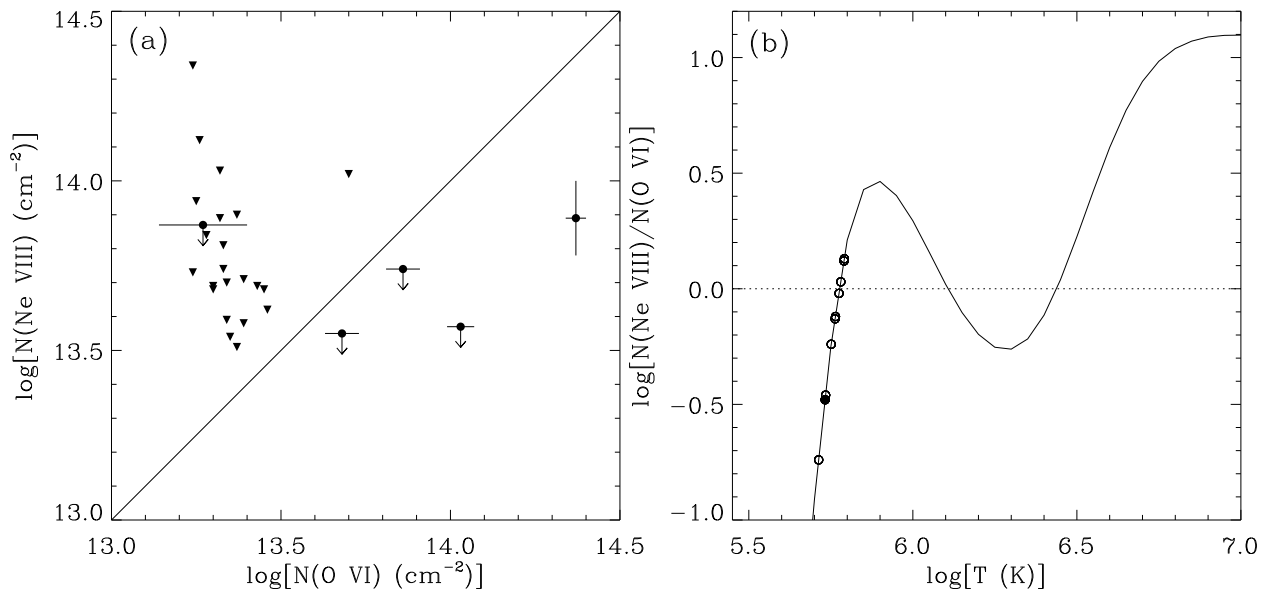}
\caption{(a) Logarithmic column densities of \neviii\ against \ovi\ observed toward HE\,0226--4110. Upside down triangles
represent data that are  3$\sigma$ upper limits for $N$(\ovi) and $N$(\neviii). 
The solid straight line  corresponds to a 1:1 relationship. (b) The solid curve represents
the relationship between the ratio of \neviii\ to \ovi\ and the temperature in a CIE model assuming a solar
relative abundance from \citet{asplund04} 
between Ne and O. The circles are observed column densities for which $N$(\ovi) is measured
and $N$(\neviii) is a measure (full circle) or a 3$\sigma$ upper limit (open circles). These data have
been obtained along the lines of sight to HE\,0226--4110, PKS\,0405--123, and 
PG\,1259+593. These data are put on the CIE curve. 
This figure shows that, if CIE applies {\em and} if the relative abundances are solar,
the WHIM  $\log T \sim  5.7$--5.8 has only been detected in one of ten \ovi\ systems. Most
\ovi\ systems trace gas with $T<5\times 10^5$ K based on the broadening of the lines. 
\label{neviiifig}}
\end{figure}

In Fig.~\ref{neviiifig}a we show the column density limits and measurement for \neviii\ against
those of \ovi\ for the HE\,0226--4110 line of sight. The upper limits on $N$(\neviii) were typically 
integrated over a velocity range of 80--100 \km, which corresponds to 
$T \sim (0.6-1.1)\times 10^6$ K if the profile is purely thermally broadened. 
Fig.~\ref{neviiifig}b shows the relationship of the 
ratio of the \neviii\ to \ovi\ column with the temperature for the CIE 
model of \citet{sutherland93} assuming a  solar relative abundance.
The circles overplotted on this curve are obtained from the measurements
of \ovi\ and measurements or upper limits on \neviii\ obtained toward 
HE\,0226--4110,  PKS\,0405--123 \citep{prochaska04,williger05},
and PG1259+594 \citep{richter04} lines of sight.   These data  cannot
fit the curve at higher temperatures because the observed 
\ovi\ broadening always implies $T<10^6$ K, except for the system at $z = 0.42660$. For this system the
limit on \neviii\ is too small. If CIE applies along with solar Ne/O, all these data
imply $T < 6 \times 10^5$ K for the \ovi\ systems.  Fig.~\ref{novib} shows that most of the detected
\ovi\ systems have column densities $>13.5$ dex. Hence, the present observations
should be sufficiently sensitive to detect \neviii\ systems with $T\sim 10^6$ K since most 
of our $3\sigma$ limits are less than 13.8 dex (see Fig.~\ref{neviiifig}a). 
We note that even though at higher temperatures $N($\neviii$) \gg N($\ovi$)$,
the \neviii\ profile would be broadened
beyond detection with the current S/N observations.  For the \ovi\ 
system at $z = 0.42660$, the broadening of the \ovi\ absorption favors a 
high temperature, but no \neviii\ is detected; 
since the \ovi\ system can be photoionized and non-thermally broadened, a cooler temperature is possible.
 
Solar reference abundances have changed quite significantly over the
recent years for C, O, and Ne. Since \citet{savage05} produced the summary of recent solar
abundance revisions (see their Table~5), recent helioseismological observations have been used 
to argue that the Ne abundance given by \citet{asplund04} is 
too low to be consistent with the solar interior \citep{bahcall05}. 
Recently, \citet{drake05} derived  Ne/O
abundances from {\em Chandra}\ X-ray spectra of solar-type stars and
they found Ne/O abundances that are 2.7 times larger than the
values reported by \citet{asplund04}, which alleviates the helioseismology problem.
But if the Ne abundance is $-3.73$ instead of $-4.16$, it would further complicate 
the explanation for the low frequency of occurrence of \neviii\ systems. 
It is still unclear what the definitive value is for the solar Ne abundance, since 
recent re-analysis of solar spectra still yield a low value \citep[e.g.,][]{young05}.

The WHIM at $T \ga 10^6$ K (where the bulk of the baryons should be formed, according
to simulations) remains still to be discovered with EUV--FUV metal-line observations. 
So far only the X-ray detections of \ovii\  in
nearby systems at $z=0.011,0.027$ discussed by 
\citet{nicastro05} imply temperatures of a few $\times 10^6$ K. 
The non-detection of hot systems with $T > 7\times 10^5$ K via \neviii\  may have several explanations
and implications. If the \ovi\ systems are mainly photoionized it is not
surprising to find very few containing \neviii. Assuming that the numerical simulations
of the WHIM are correct,  
the bulk of the WHIM may have a lower metallicity than the gas traced by hot \ovi\ absorbers.
Broad Ly$\alpha$ absorbers which trace the WHIM appear in some cases to have a 0.01 solar
abundance (see \S\ref{coll}). The relative abundances may not be always solar. 
Low metallicity environments are known to have relative abundances different than those in the Sun. 
CIE models may not be a valid representation of the distribution of ions in hot gas in the WHIM, 
but we note that the non-equilibrium cooling flow models by \citet{heckman02} would
generally also predict too much \neviii\ (see Table~\ref{t5}). Additional
and more sensitive searches for \neviii\ would be valuable for better statistics and a better
understanding of the \neviii\ systems, and their frequency of occurrence in 
the IGM. 

\subsection{Metallicity of the Low Redshift Metal-Systems}\label{metallicity}

Metallicity is a quantity that is fundamental in obtaining estimates of $\Omega_{\rm b}$ from 
metal-line absorbers  and following the chemical enrichment of the IGM. 
A metallicity of 0.1 solar is generally adopted for 
estimating $\Omega_{\rm b}({\mbox \ovi})$ \citep[e.g.,][]{tripp00b,danforth05}, but 
since $\Omega_{\rm b}\propto Z^{-1}$, a change in the adopted metallicity
can significantly alter the estimated $\Omega_{\rm b}$. 
Table~\ref{t4a} provides recent metallicity estimates
via photoionization models of \ovi\ systems in the low redshift IGM. If 
these systems are primarily photoionized, they are not 
tracers of the WHIM. But abundance measurements in collisionally
gas are generally not reliable. Combining the results in Table~6, 
we find a median abundance for the photoionized \ovi\ systems 
of $-0.5$ dex solar with a large scatter of $\pm 0.5$ dex around this value.
Only 4 of 13 systems appear to be either at solar abundance or
less than 1/10 solar. If those systems are not taken into account, 
the metallicity is $-0.45 \pm 0.10$ dex. 
If the metallicity of the photoionized \ovi\ systems
reflects the metallicity of the collisionally ionized \ovi\ systems, 
$\Omega_{\rm b}$ derived via \ovi\ would decrease by a factor $\sim3$ since
$\Omega_{\rm b}\propto Z^{-1}$.

\section{Summary}\label{sum}
We present the analysis of the intervening absorption and the interpretation of the metal-line systems
in the {\fuse}\ and STIS FUV spectra of the QSO HE\,0226--4110 ($z_{\rm em} = 0.495$).
Due to the low fraction of Galactic ISM molecular hydrogen along this line of sight, 
HE\,0226--4110 provides an excellent opportunity to search for metals associated with the 
Ly$\alpha$ forest. For each Ly$\alpha$ absorber, we systematically search for 
\ciii, \oiii, \oiv, \ovi, and \neviii. We also search for \ovi\ and \neviii\
doublets not associated with \hi. For each metal-line system detected, we also search for any
other metals that may be present. The richest intervening metal-line system along the
HE\,0226--4110 sight line is at $z=0.20701$. This system was fully discussed by \citet{savage05}. 
We examine the ionic ratios to constrain the ionization mechanisms, the metallicity, and physical conditions of the absorbers
using single-phase photoionization and collisional ionization models. 
We discuss our results in conjunction with analyses of the metal-line systems
observed in several other low redshift QSO spectra of similar quality. The main results are as follows: 
\begin{enumerate}

\item We detect 4 \ovi\ absorbers with rest frame equivalent widths $W_\lambda\ga 50$ m\AA.  
A tentative detection at $z=0.42660$
is also reported but this system could be misidentified because Ly$\alpha$ cannot be accessed.  
The number of  intervening \ovi\ systems with  $W_\lambda \ga 50$ m\AA\ per unit redshift
is $d{\mathcal N}({\mbox \ovi})/dz \approx 11$ along the HE\,0226--4110 line of sight. 
For 4 of the 5 \ovi\ systems other ions (such as \ciii, \civ, \oiii, \oiv) are detected.

\item One \ovi\ system at $z = 0.20701$ cannot be explained by photoionization by the UV background
(see Savage et al. 2005). For the other 4 \ovi\ systems, photoionization can reproduce the
observed column densities. However, for these systems, collisional ionization may also 
be the origin of the \ovi.  We note that if photoionization applies, 
the broadening of the metal lines must be mostly non-thermal, but the \hi\ broadening is mostly
thermal. 

\item Oxygen is particularly useful to study the ionization and physical conditions in the IGM because 
a wide range of ion states are accessible at  FUV and EUV 
restframe wavelengths, including \oi, \oii, \oiii, \oiv, \ov, 
and \ovi. Our results imply that $d{\mathcal N}({\mbox \oii})/dz \ll  d{\mathcal N}({\mbox \oiii})/dz 
\le d{\mathcal N}({\mbox \oiv})/dz \approx d{\mathcal N}({\mbox \ovi})/dz $. 
The low redshift IGM is more highly ionized than weakly ionized
since the transitions for the different oxygen ions have similar values of $\lambda f$.

\item Following \citet{prochaska04},   we confirm with a larger sample
that the photoionized metal-line systems have a decreasing ionization parameter
with increasing \hi\ column density.
The \ovi\ systems that can be explained by both photoionization and collisional 
ionization follow this relationship. This implies
that the systems with photoionization/CIE degeneracy may likely be photoionized. 

\item Combining our results with those toward 3 other sight lines, we 
show that there is a general increase of $N$(\ovi) with increasing $b$(\ovi) but with 
a large scatter. The observed distribution of $N$(\ovi) and $b$(\ovi) can 
be reproduced by cooling flow models computed by \citet{heckman02}, but these
models fail to reproduce the observed ionic ratios. 
 
\item  Combining results for several QSOs 
we find that the photoionized \ovi\ systems in the low redshift IGM 
have a median abundance of 0.3 solar, about 3 times the metallicity
typically used to derive $\Omega_b$ in the warm-hot IGM gas.
 
\item Along the path to HE\,0226--4110, there is one detection 
of \neviii\ at $z=0.20701$ that implies a gas temperature of $5.4\times 10^5$ K. 
We did not detect other \neviii\  systems  
although our sensitivity should have allowed the detection of \neviii\ in \ovi\ systems
at $T \sim (0.6-1.3 )\times 10^6$ K if CIE applies and Ne/O is solar. 
Since the bulk of the warm-hot ionized medium (WHIM) is believed to be at temperatures $T> 10^6 $ K,
the hot part of the WHIM remains to be discovered with FUV--EUV metal-line transitions.

\item This work shows that the origins 
of the \ovi\ absorption (and the metal-line systems in general) 
are complex and cause several uncertainties in attempts to estimate 
$\Omega_{\rm b}$ from \ovi. In particular the \ovi\ ionization mechanism in several cases 
is indeterminate. High S/N 
FUV spectra obtained with COS  or other future FUV instruments 
will be needed to solve several of the current observational uncertainties.

\end{enumerate}

\acknowledgments

We thank Marilyn Meade for calibrating the many {\em FUSE}\ datasets. 
The STIS observations of HE0226-4110 were obtained as part of {\it
HST}\ program 9184 with financial support from NASA grant HST
GO-9184.08-A. 
Support for this research was provided by NASA through grant 
HST-AR-10682.01-A. NL was also supported by NASA grants NNG04GD885G and NAG5-13687.
BPW was supported by NASA grants LTSA NAG5-9179 and ADP NNG04GD885G.
TMT was supported in part by NASA LTSA grant NNG04GG73G.
This research has made use of the NASA
Astrophysics Data System Abstract Service and the SIMBAD database,
operated at CDS, Strasbourg, France.

\begin{deluxetable}{lcc}
\tabcolsep=3pt
%\rotate
\tablecolumns{3}
\tablewidth{0pt} 
\tabletypesize{\footnotesize}
%\tabletypesize{\scriptsize}
\tablecaption{STIS Observation of HE\,0226--4110 \label{t1a}} 
\tablehead{\colhead{ID}    &  \colhead{Date}  &  \colhead{Exp. Time} \\ 
\colhead{}  &\colhead{}&\colhead{(ks)}}
\startdata
O6E107010	&	2002-12-25 	&  2.1 \\
O6E107020	&	2002-12-26 	&  3.0 \\
O6E107030	&	2002-12-26 	&  3.0 \\
O6E108010	&	2002-12-26 	&  2.1 \\
O6E108020	&	2002-12-26 	&  3.0 \\
O6E108030	&	2002-12-26 	&  3.0 \\
O6E109010	&	2002-12-26 	&  2.1 \\
O6E109020	&	2002-12-27 	&  3.0 \\
O6E109030	&	2002-12-27 	&  3.0 \\
O6E110010	&	2002-12-29 	&  2.1 \\
O6E110020	&	2002-12-29 	&  3.0 \\
O6E110030	&	2002-12-29 	&  3.0 \\
O6E111010	&	2002-12-31 	&  2.1 \\
O6E111020	&	2002-12-31 	&  3.0 \\
O6E111030	&	2003-01-01 	&  3.0 \\
O6E111040	&	2003-01-01 	&  3.0 \\
		&			& Total\,$ =43.5 $  \\
\enddata		
\end{deluxetable}

\begin{deluxetable}{lcccccccccc}
\tabcolsep=3pt
%\rotate
\tablecolumns{11}
\tablewidth{0pt} 
\tabletypesize{\footnotesize}
%\rotate
\tabletypesize{\scriptsize}
\tablecaption{{\fuse}\ Observations of HE\,0226--4110 \label{t1}} 
\tablehead{\colhead{ID}    &  \colhead{Date} &  \colhead{CALFUSE}  &  \colhead{Mode}   &  \colhead{Flux(1A)} &  \colhead{Flux(2B)} &
  \colhead{Flux(Comb.)} &  \colhead{$t_{\rm exp}($1A)$^a$}&  \colhead{$t_{\rm exp}($1B)$^a$} &  \colhead{$t_{\rm exp}($2A)$^a$} &  \colhead{$t_{\rm exp}($2B)$^a$} \\ 
\colhead{} &\colhead{}  &\colhead{} &\colhead{} &\colhead{$10^{-14}$ cgs}&\colhead{$10^{-14}$ cgs}&\colhead{$10^{-14}$ cgs}&\colhead{(ks)}&\colhead{(ks)}&\colhead{(ks)}&\colhead{(ks)}}
\startdata
P2071301	& 	2000-12-12  & v2.1.6	&	        & 2.60  &  2.60  &  2.59   & 11.0  (8.4) & 11.0  (8.4) & 11.0  (8.4) & 11.1   (8.4)	  \\	  
P1019101	&	2001-10-03  & v2.1.6	&     FP-SPLIT  & 2.57  &  2.73  &  2.62   & 50.3 (37.9) & 50.1 (37.9) & 50.0 (37.9) & 49.9  (38.0)   \\ 
P1019102	&	2002-11-15  & v2.4.0	&     FP-SPLIT  & 2.44  &  2.06  &  2.30   & 14.5 (11.1) & 14.5 (11.0) & 14.5 (11.0) & 14.5  (11.1)   \\ 
P1019103	&	2002-11-16  & v2.4.0	&     FP-SPLIT  & 2.45  &  2.19  &  2.35   & 18.9 (14.7) & 18.9 (14.7) & 18.9 (14.7) & 18.9  (14.7)   \\ 
P1019104	&	2002-11-17  & v2.4.0	&     FP-SPLIT  & 2.57  &  2.39  &  2.50   & 18.1 (10.5) & 18.1 (10.5) & 18.1 (10.5) & 18.1  (10.5)   \\ 
D0270101	&	2003-09-01  & v2.4.0	&	        & 3.06  &  2.80  &  2.97   & 23.9 (12.5) & 24.7 (12.6) & 23.9 (12.4) & 23.9  (12.5)   \\ 
D0270102	&	2003-09-03  & v2.4.0	&	        & 3.08  &  2.98  &  3.05   & 39.9 (19.2) & 39.4 (19.2) & 37.6 (19.1) & 37.3  (19.1)   \\ 
D0270103	&	2003-10-31  & v2.4.0	&	        & 3.25  &  3.16  &  3.22   & 16.8 (11.9) & 16.8 (12.0) & 17.0 (12.0) & 17.1  (12.0)   \\
\enddata		
\tablecomments{$a:$ The first number is the total exposure time. The night-only exposure time
is in parenthesis. 
}
\end{deluxetable}

\begin{deluxetable}{lcccccccc}
\tabcolsep=2pt
%\rotate
\tablecolumns{9}
\tablewidth{0pt} 
\tabletypesize{\footnotesize}
%\tabletypesize{\scriptsize}
\tablecaption{Properties of the Intervening Systems toward HE\,0226--4110 \label{t2}} 
\tablehead{\colhead{Species}    &  \colhead{$\lambda_{\rm obs}$}  &  \colhead{$W_\lambda$} &\colhead{$v$}   & \colhead{$b$}   &  \colhead{$\log N$} & \colhead{Method}& \colhead{$[-v,+v]$}& \colhead{Instrument} \\ 
\colhead{}  &\colhead{(\AA)}&\colhead{(m\AA)} & \colhead{(\km)}&\colhead{(\km)}   &\colhead{(dex)} &\colhead{} &\colhead{(\km)}& \colhead{}\\
\colhead{(1)}& \colhead{(2)} & \colhead{(3)}& \colhead{(4)}& \colhead{(5)}&\colhead{(6)}&\colhead{(7)}&\colhead{(8)}&\colhead{(9)}}
\startdata
\cutinhead{$z = 0.01746^a $}
\hi\ $\lambda$1215	       	   	&  1236.944	&  $79.4 \pm12.1$	&  $+1.2 \pm 2.9$       &  $22.7 \pm  3.2$	     &  $13.28 \pm 0.07$	       & AOD 	    &	$[-40,40]$	& STIS       \\
\hi\ $\lambda$$\lambda$1215,1025       	&  1236.944	&  \nodata		&  $+0.0 \pm 2.2 $      &  $17.9 \pm 3.9$ 	     &  $13.22 \pm 0.06$	       & FIT	    &	\nodata		& STIS,LiF\,2B       \\
\cii\ $\lambda$1036		   	& 1054.473   	&  $<23.3 $		&  \nodata		&  \nodata		     &  $< 13.32		  $    & 3$\sigma$  &	$[-40,40]$	& LiF\,1A	   \\
\ciii\ $\lambda$977		   	& 994.118	&  $< 65.4 $		&  \nodata		&  \nodata		     &  $< 13.01		  $    & 3$\sigma$  &	$[-40,40] $	& LiF\,1A	   \\
\civ\ $\lambda$1548	 	   	& 1575.289    	&  $41.3 \pm 13.9 $     &  $+0.4 \pm  4.8 $     & $ 10.0 :       $	     &  $13.14 \pm\,^{0.12}_{0.18}$    & AOD	    &	$[-30,30]$	& STIS  	\\
\civ\ $\lambda$1550	 	   	& 1577.909    	&  $36.8 \pm 15.1 $     &  $+9.6 \pm  5.8 $     & $ 21.0 \pm5.5 $	     &  $13.34 \pm\,^{0.14}_{0.22}$    & AOD	    &	$[-30,30]$	& STIS  	\\
\civ\ $\lambda$$\lambda$1548,1550  	&   \nodata	&  \nodata		&  $-0.4 \pm  2.5 $     & $10.0 \pm 6.8  $	     &  $13.13 \pm 0.11 	  $    & FIT	    &	\nodata		& STIS  	 \\
\nv\ $\lambda$1242		   	& 1264.554    	&  $< 32.6 $		&  \nodata		&  \nodata		     &  $< 13.48		  $    & 3$\sigma$  &	$[-40,40] $	& STIS  		\\
\enddata
\end{deluxetable}
 
\begin{deluxetable}{lccccccc}
\tabcolsep=3pt
%\rotate
\tablecolumns{8}
\tablewidth{0pt} 
\tabletypesize{\footnotesize}
%\tabletypesize{\scriptsize}
\tablecaption{IGM Absorbers toward HE\,0226--4110 \label{t3}} 
\tablehead{\colhead{$z$}    &  \colhead{$\log N($\hi)}  &  \colhead{$b($\hi)} &\colhead{$\log N($\ciii)}   & \colhead{$\log N($\oiii)}   &  \colhead{$\log N($\oiv)} & \colhead{$\log N($\ovi)}& \colhead{$\log N($\neviii)}}
\startdata
0.01746 	& $13.22 \pm 0.06$   & $17.9 \pm 4.3$  &  $< 13.01		    $  &     n.a.		&	n.a.	& $13.60 \pm 0.10 	$ & n.a. \\
0.02679 	& $13.22 \pm 0.08$   & $41.6 \pm11.0$  &  $< 12.78		    $  &     n.a.		&	n.a.	& $< 13.34		$ & n.a. \\
0.04121 	& $12.82 \pm 0.14$   & $23.6 \pm18.6$  &  $< 12.43		    $  &    n.a.		&	n.a.	& $< 13.50		$ & n.a. \\
0.04535 	& $12.71 \pm 0.13$   & $16.2 \pm12.8$  &  $< 12.44		    $  &     n.a.		&	n.a.	& $< 13.26		$ & n.a. \\
0.04609 	& $13.66 \pm 0.03$   & $25.0 \pm 2.1$  &  $< 12.64		    $  &     n.a.		&	n.a.	& $< 13.37		$ & n.a. \\
0.06015 	& $13.19 \pm 0.06$   & $35.5 \pm 7.6$  &  $< 12.61		    $  &     n.a.		&	n.a.	& $< 13.30		$ & n.a. \\
0.06083 	& $14.65 \pm 0.02$   & $44.5 \pm 1.0$  &  \nodata		       &     n.a.		&	n.a.	& $< 13.33		$ & n.a. \\
0.07023 	& $13.81 \pm 0.11$   & $26.0 \pm12.4$  &  $< 12.41		    $  &     n.a.		&	n.a.	& $< 13.24		$ & n.a. \\
\enddata		
\tablecomments{``...": species cannot be observed because of blending; ``n.a.": species is not observable at this redshift. 
``!": Possible misidentification, see \S\ref{misid} for more details.  
$^a$The total column densities are presented, but the structure is more complex (see Savage et al. 2005). 
$^b$These three system correspond to the 3 absorption lines observed in the system denoted 
$z \simeq 0.27155$ in Table~\ref{t2}.
$^c$The limit is based on the Ly$\beta$ line (no detection), the 
corresponding Ly$\alpha$ line is redward of the available STIS wavelength range. 
}
\end{deluxetable}

\begin{deluxetable}{lcccccccc}
\tabcolsep=3pt
%\rotate
\tablecolumns{9}
\tablewidth{0pt} 
\tabletypesize{\footnotesize}
%\tabletypesize{\scriptsize}
\tablecaption{Summary of Physical Properties of the \ovi\ Systems toward HE\,0226--4110 \label{t4}} 
\tablehead{\colhead{$z$}    &  \colhead{$\log N($\hi)}  &  \colhead{$b($\hi)} &  \colhead{$\log N($\ovi)} & 
\colhead{Ionization$^a$}& \colhead{$[{\rm O/H}]^b$} &  \colhead{$\log N($H$)$$^b$} &  \colhead{$\log U^b$}&  \colhead{$L^b$}\\
\colhead{}    &  \colhead{dex}  &  \colhead{\km} &  \colhead{dex} & \colhead{\km} & \colhead{dex} &  \colhead{dex} &  \colhead{dex} &  \colhead{kpc}}
\startdata
0.01746 	& $13.22  $   & $19.8  $  & $13.80   $ &  Photo      &  $0.0 $  	& $17.50$          & $-1.10$    & $17$  	  \\
0.20701         & $15.28^c$   & $39.9^c$  & $ 14.37  $ &  Coll      &  $-0.5$ 	   	& $19.92$         & \nodata    & $<10^3$  	\\
0.34034         & $13.47  $   & $31.2  $  & $13.86   $ & Photo/Coll  & $-0.3 $		& $18.22$         & $-1.00$     & $40$  	\\
0.35523         & $13.60  $   & $27.1  $  & $13.68   $ & Photo/Coll  &$-0.55$		& $18.16$          & $-0.99$    & $40$  	  \\
0.42660$^!$     & $< 13.55$    & \nodata  & $ 14.03  $ & Photo/Coll  & \nodata 		& \nodata         & \nodata     & \nodata  	\\
\enddata		
\tablecomments{$^a$This column lists the ionization mechanism for \ovi: photoionization (photo), collisional ionization (coll).
``Photo/Coll" indicates that both ionization mechanisms could explain the observed \ovi\ column densities. 
$^b$The metallicity, ionization parameter, H column density, and path length of the absorber are from the CLOUDY photoionization models,
except for the \ovi\ system  at $z=0.20701$ which are from CIE (see Savage et al. 2005 for more details).
$^c$ Note this is the total \hi\ column density measured in the photoionized system that does not contain \ovi. The estimated \hi\ parameters
associated to the \ovi\ system are $\log N($\hi$)=13.70$ and $b=97$ \km\ (see Savage et al. 2005).
``!": Possible misidentification, see \S\ref{misid} for more details.  
}
\end{deluxetable}

\begin{deluxetable}{lccc}
\tabcolsep=3pt
%\rotate
\tablecolumns{4}
\tablewidth{0pt} 
\tabletypesize{\footnotesize}
\tablecaption{Summary of Photoionization Parameters and Metallicity in the low-$z$ IGM for the \ovi\ Systems \label{t4a}} 
\tablehead{\colhead{$z$}    &  \colhead{$\log N($\hi)}  &  \colhead{$\log Z/Z_\sun$} &  \colhead{$\log U$}} 
\startdata
\cutinhead{HE\,0226--4110}
0.01746	 		    &	    13.22  & $+0.0$	&$-1.1$      \\ %&    17.5    \\
0.34034	 		    &	    13.47  & $-0.3$	&$-1.0$      \\% &   18.2\\
0.35523	 		    &	    13.60  & $-0.6$	&$-1.0$      \\ %&   18.3\\
\cutinhead{PG\,0953+415}    	 	 	       
0.06807  		    &	    14.39  &  $-0.4$	& $-1.4$     \\ % &  18.5\\
\cutinhead{PG\,1116+215}    	 	 	       
0.13847	 		    &	    16.20  &  $-0.5$	& $-2.5$      \\%&   19.1\\
\cutinhead{PG\,1259+593}    	 	 	       
0.21949	 		    &	    15.25 &  $-1.3$	& $-1.5$     \\%&   19.5\\
0.29236		 	    &	    14.50 &  $-0.5$	& $-1.7$     \\%&   17.9\\
\cutinhead{PKS\,0405--123}  	 	 	       
0.09180		 	    &	    14.52 & $>-1.4$	& $>-1.5$    \\%&   18.8\\
0.09658	 		    &	    14.65 &  $-1.5$	& $-1.2$     \\%&   19.2    \\
0.16710	 		    &	    16.45 &  $-0.3$	& $-2.9$     \\%&   18.8    \\
0.36080		 	    &	    15.12 &  $>-0.7$	& $-2.0$     \\ %&   18.6    \\
0.36332 	 	    &	    13.43 &  $0.0$	& $-1.4$      \\ % 17.4\\
0.49510		 	    &	    14.39 &  $>-0.3$	& $-1.3$     \\ %&   18.5    \\
\enddata		
\tablerefs{
HE\,0226--4110: this work; PG\,0953+415: Savage et al. 2002; PG\,1116+215: Sembach et al. 2004;
PG\,1259+593: Richter et al. 2004; PKS\,0405--123: Prochaska et al. 2004.
}
\end{deluxetable}

\begin{deluxetable}{lccc}
\tabcolsep=3pt
%\rotate
\tablecolumns{4}
\tablewidth{0pt} 
\tabletypesize{\footnotesize}
%\tabletypesize{\scriptsize}
\tablecaption{Comparison of Observed Logarithmic Column density ratios with those from 
Cooling Flow models \label{t5}} 
\tablehead{\colhead{$z$}    &  \colhead{\oiv/\ovi}  &  \colhead{\neviii/\ovi}  &  \colhead{\svi/\ovi} }
\startdata
\cutinhead{Isobaric  Model}
		&  $-0.6$	& $+0.1$	& $-2.8$	\\
\cutinhead{Isochoric Model}
		&  $0.0$	& $-0.3$	& $-2.4$	\\
\cutinhead{HE\,0226--4110}
0.20701		& $>0.11$	& $-0.48$	& $-1.59$	\\
0.34034		& $-0.03$	& $<-0.55$	& $<-1.13$	\\
0.35523		& $-0.04$	& $<-0.13$	& $<-0.88$	\\
0.42660		& $<-0.49$	& $<-0.46$	& $<-1.18$	\\
\cutinhead{PG\,1259+593}
0.21949		& $-0.20$	& $<-0.10$	& \nodata 	\\
\cutinhead{PKS\,0405--123}
0.16701		& \nodata	& \nodata	& $-1.90$	\\
0.18292		& $<-0.57$	& \nodata	& \nodata 	\\
0.36335		& $+0.21$	& $<-0.03$	& \nodata 	\\
0.49512		& $>-0.15$	& $(<)-0.74$	& $<-1.87$	\\
\enddata		
\tablecomments{The calculations of the cooling columns at constant
pressure (isobaric) and density  (isochoric) are from \citet{heckman02}. 
The measurements are from: HE\,0226--4110, this paper; PKS\,0405--123, \citet{prochaska04}
and Williger et al. (2005); PG\,1259+593, Richter et al. (2004). 
}
\end{deluxetable}

\begin{deluxetable}{lccc}
\tabcolsep=3pt
%\rotate
\tablecolumns{4}
\tablewidth{0pt} 
\tabletypesize{\footnotesize}
%\tabletypesize{\scriptsize}
\tablecaption{EUV--FUV Oxygen Transitions \label{t6}} 
\tablehead{\colhead{Ion}    &  \colhead{$\lambda$ (\AA)}  &  \colhead{$\log \lambda f $}  &  \colhead{$ z$} \\
\colhead{(1)}    &  \colhead{(2)}  &  \colhead{(3)}  &  \colhead{(4)} 
}
\startdata
\oi\ 	&  971.738 &  1.10 &  $>0	  $ \\
\oi\ 	&  988.773 &  1.65 &  $>0	  $ \\
\oi\ 	& 1039.230 &  0.97 &  $>0	  $ \\
\oi\ 	& 1302.169 &  1.82 &  $> 0	  $ \\
\oii\ 	&  832.757 &  1.57 &  $>0.105	   $ \\
\oii\ 	&  833.329 &  1.87 &  $>0.104	   $ \\
\oii\ 	&  834.466 &  2.04 &  $>0.103	   $ \\
\oiii\ 	&  702.332 &  1.98 &  $>0.310	   $ \\
\oiii\ 	&  832.927 &  1.95 &  $>0.104	   $ \\
\oiv\   &  553.330 &  1.79 &  $>0.663	   $ \\
\oiv\   &  554.075 &  2.09 &  $>0.660	   $ \\
\oiv\   &  608.398 &  1.61 &  $>0.339	   $ \\
\oiv\ 	&  787.711 &  1.94 &  $>0.169	   $ \\
\ov\  	&  629.730 &  2.51 &  $>0.463	   $ \\
\ovi\ 	& 1031.926 &  2.14 &  $>0	  $ \\
\ovi\ 	& 1037.617 &  1.83 &  $>0	  $
\enddata		
\tablecomments{Column 4 shows the lower redshift at which a given transition can be observed in the FUV.  
We only list the strongest FUV \oi\ transitions since they are rarely detected in the IGM.
Atomic parameters can be found in Morton 2005 (FUV transitions) and Verner et al. 1996 (EUV transitions).
}
\end{deluxetable}

\clearpage

\clearpage
%\begin{figure}[tbp]
%\epsscale{0.9} 
\plotone{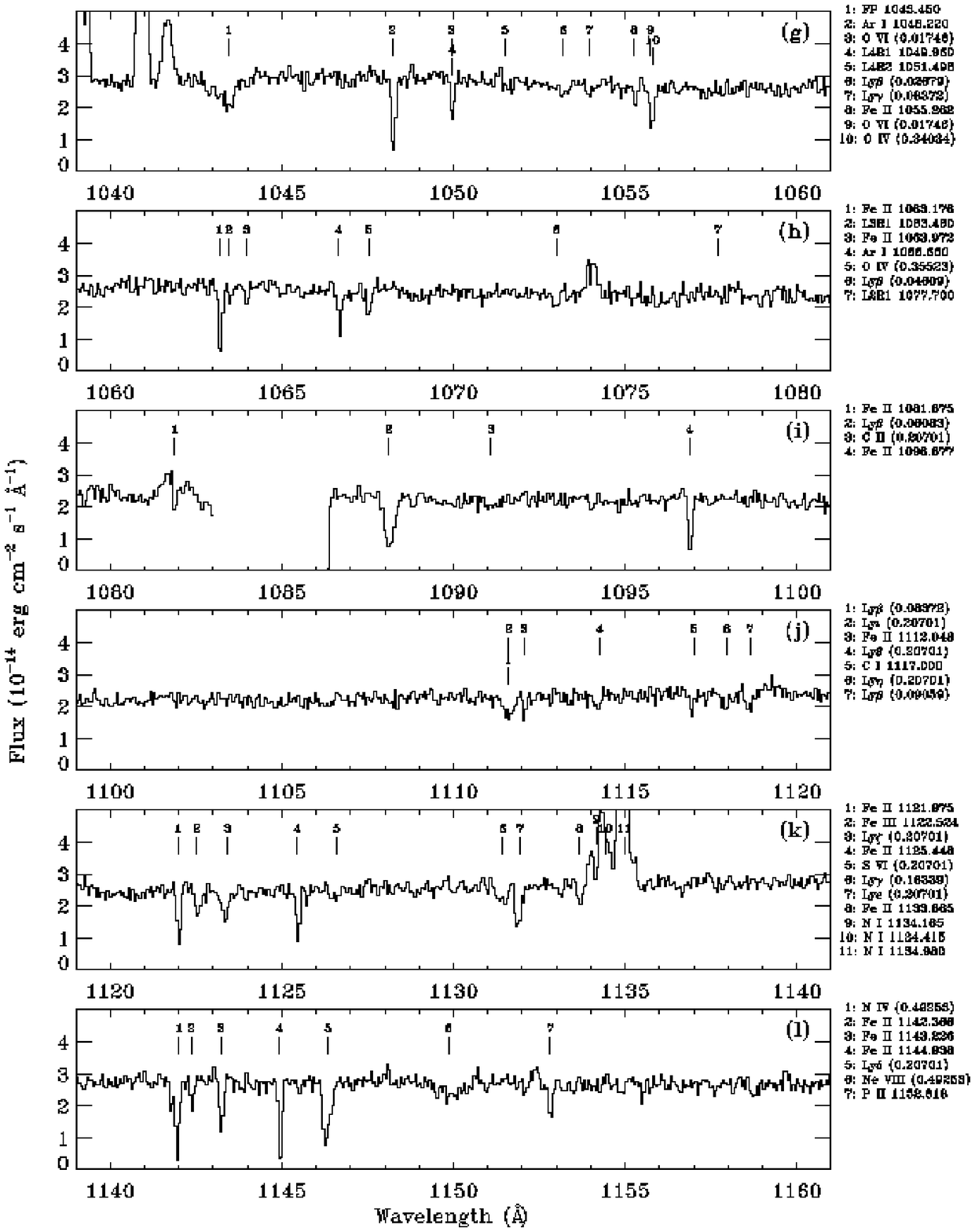}
%\figurenum{\ref{fig2}}
\centerline{Fig. 2 --- Continued.}
%\end{figure}
\clearpage
%\begin{figure}[tbp]
%\epsscale{0.9} 
\plotone{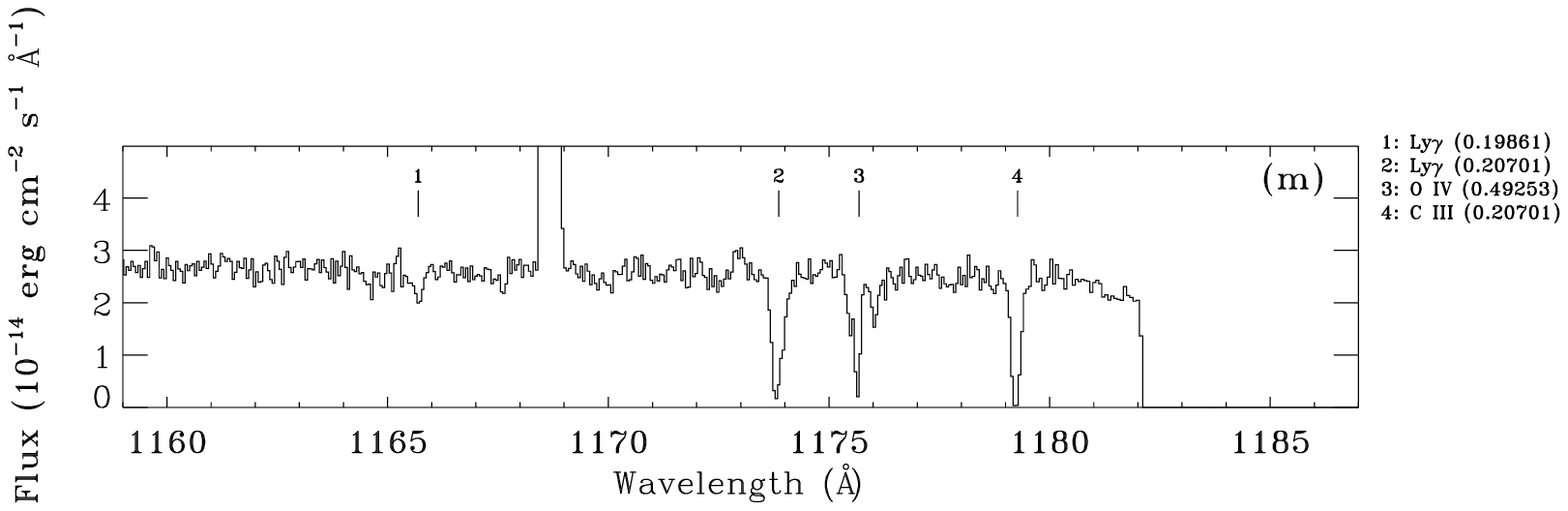}
%\figurenum{\ref{fig2}}
\centerline{Fig. 2 --- Continued.}
%\end{figure}
\clearpage

\clearpage
%\begin{figure}[tbp]
%\epsscale{0.9} 
\plotone{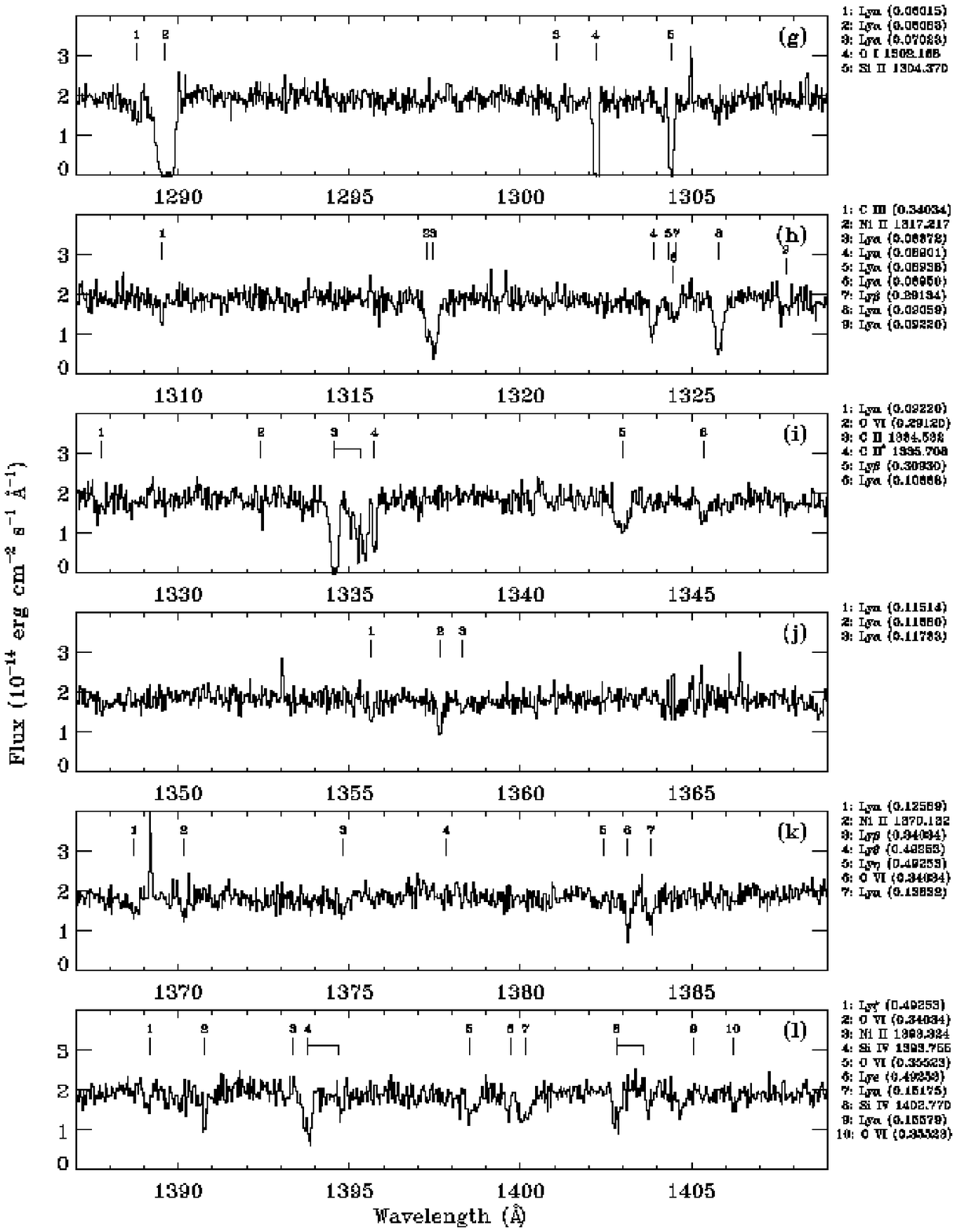}
\centerline{Fig. 3 --- Continued.}
%\figurenum{\ref{fig3}}
%\end{figure}
\clearpage
%\begin{figure}[tbp]
%\epsscale{0.9} 
\plotone{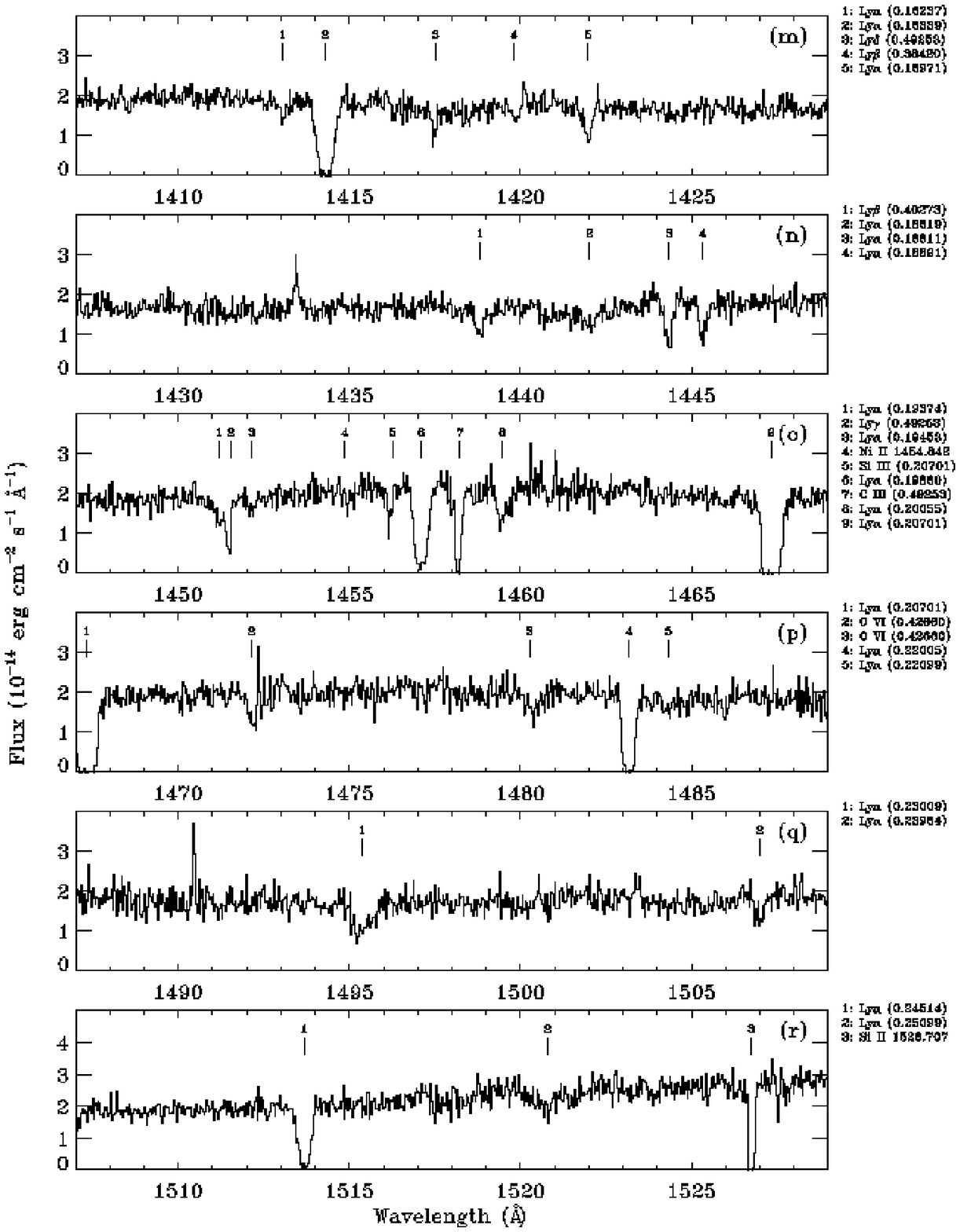}
\centerline{Fig. 3 --- Continued.}
%\figurenum{\ref{fig3}}
%\caption{continued.}
%\end{figure}
\clearpage
%\begin{figure}[tbp]
%\epsscale{0.9} 
\plotone{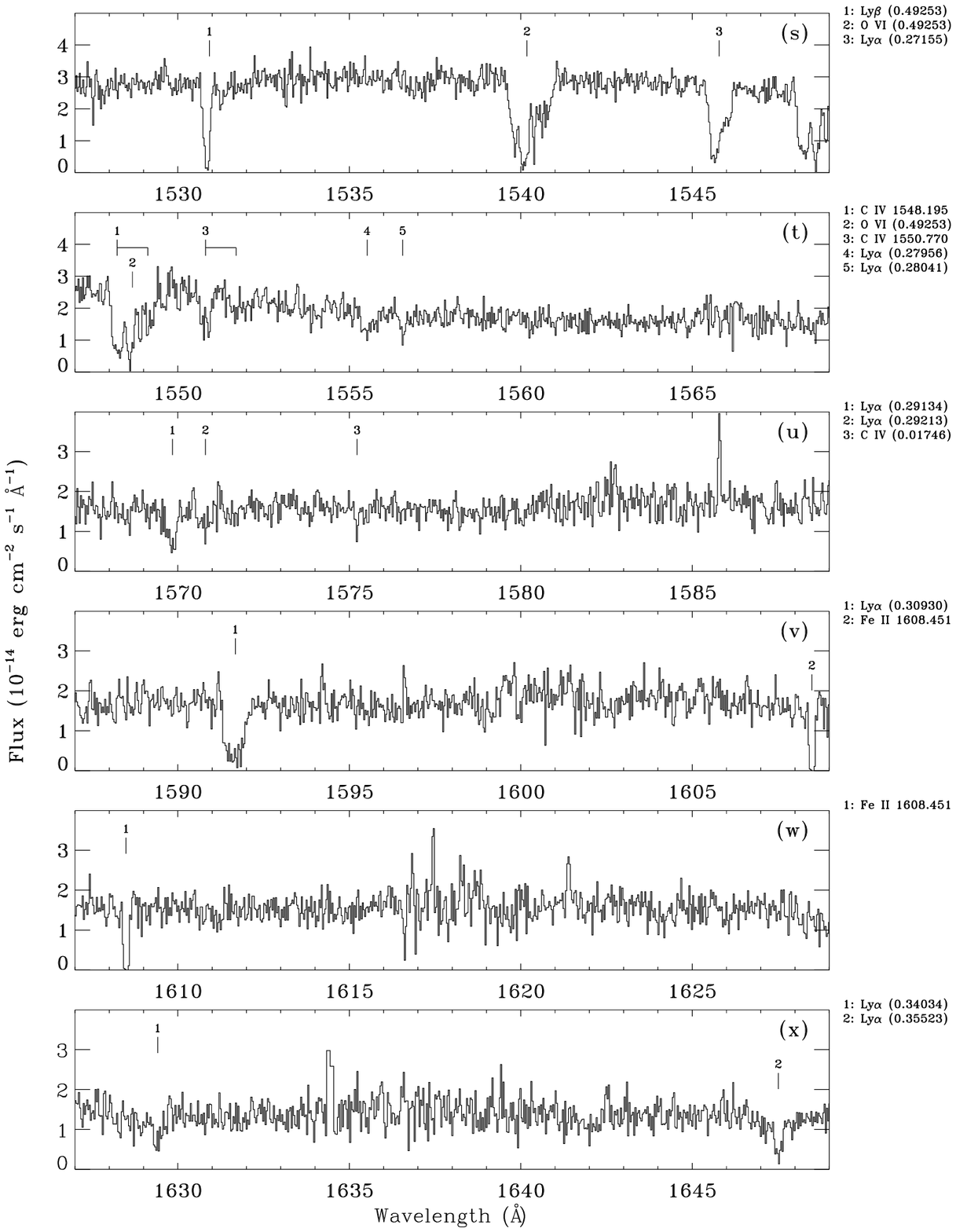}
\centerline{Fig. 3 --- Continued.}
%\figurenum{\ref{fig3}}
%\caption{continued.}
%\end{figure}
\clearpage
%\begin{figure}[tbp]
%\epsscale{0.9} 
\plotone{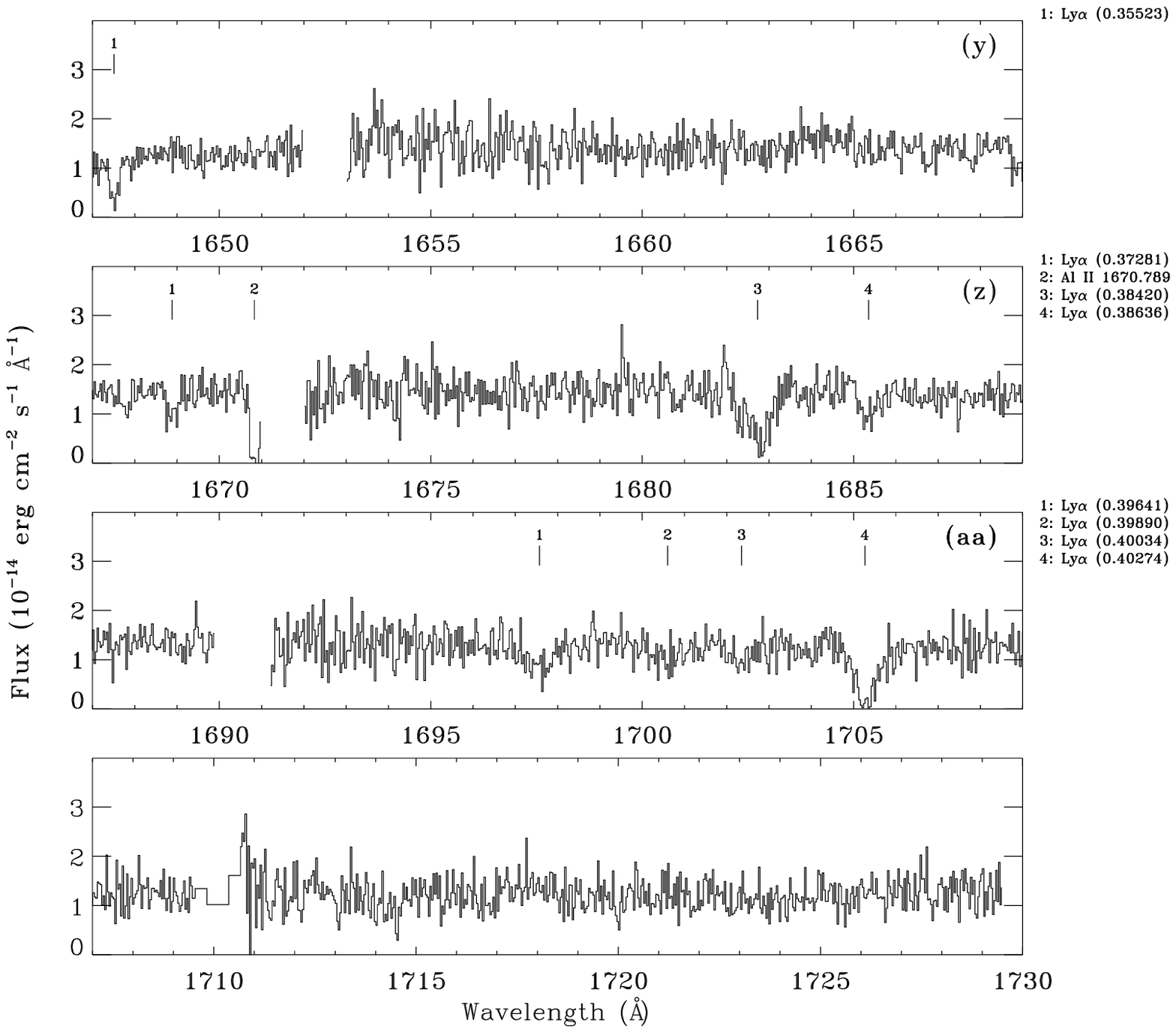}
\centerline{Fig. 3 --- Continued.}
%\figurenum{\ref{fig3}}
%\caption{continued.}
%\end{figure}
\clearpage
\clearpage
%\begin{figure}[tbp]
%\epsscale{0.9} 
\plotone{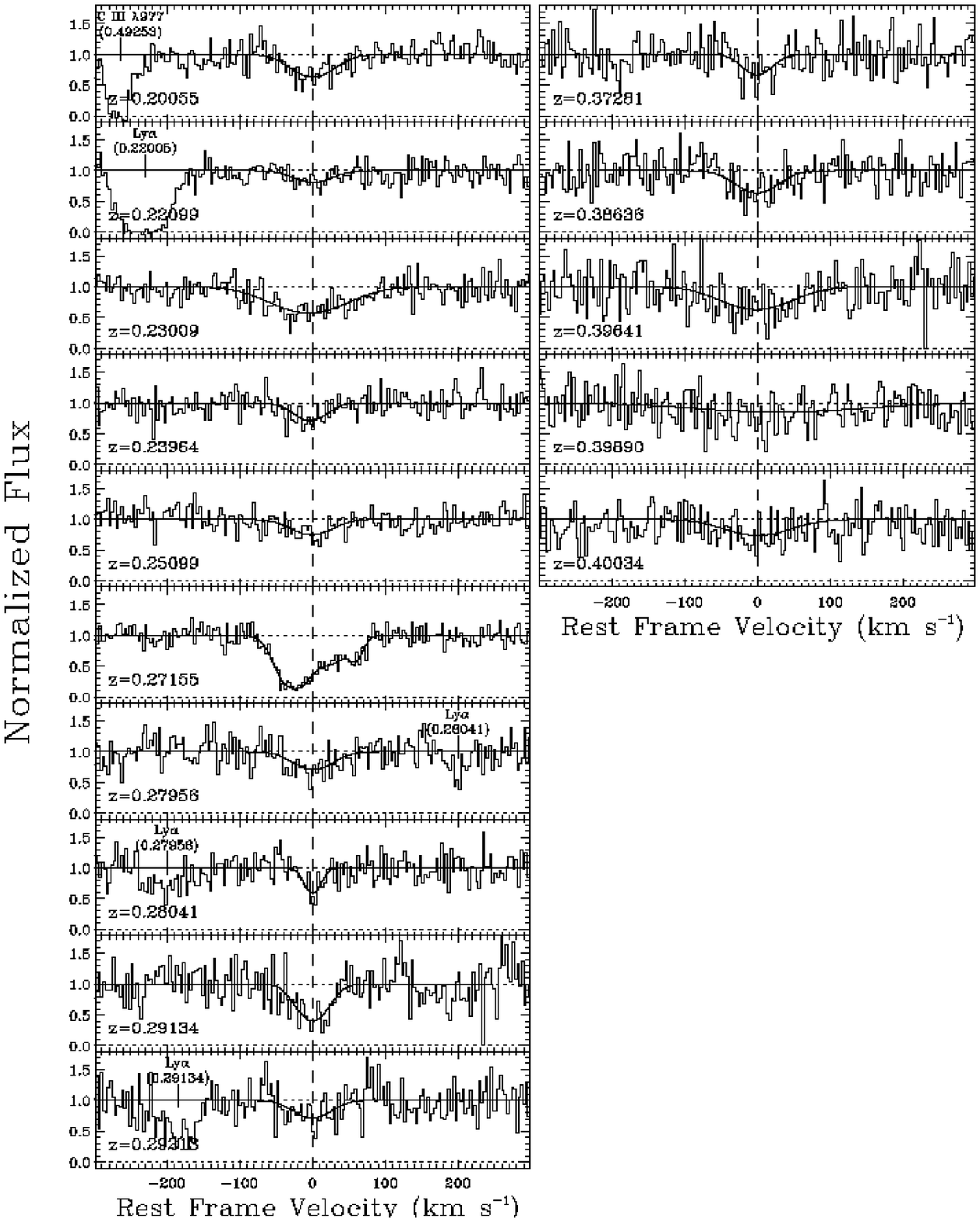}
\centerline{Fig. 4 --- Continued.}
%\figurenum{\ref{hi1215only}}
%\caption{continued.}
%\end{figure}

\clearpage

\end{document}